\documentclass[%
 aip,
 amsmath,amssymb,
 reprint,%
]{revtex4-1}

\usepackage{graphicx}
\usepackage{dcolumn}
\usepackage{bm}

\usepackage[utf8]{inputenc}
\usepackage[T1]{fontenc}
\usepackage{mathptmx}
\usepackage{etoolbox}

\usepackage{comment}

\usepackage{hyperref}
\hypersetup{
           hidelinks,
           breaklinks=true,   
           pdfusetitle=true,  
        }

\newcommand{\km}{k_\mathrm{m}}

\newcommand{\ko}{k_\mathrm{o}}

\newcommand{\om}{\omega_{\rm m}}
\newcommand{\Um}{U_{\rm m}}
\newcommand{\Umprime}{U'_{\rm m}}
\newcommand{\omo}{\omega_{\rm o}}
\newcommand{\Qm}{Q_{\rm m}}
\newcommand{\omopi}{\omega_{\rm m}/(2\pi)}

\newcommand{\Qo}{Q_{\rm o}}
\newcommand{\ooopi}{\omega_{\rm o}/(2\pi)}

\newcommand{\gom}{g_\mathrm{om}}
\newcommand{\nc}{n_\mathrm{c}}
\newcommand{\Com}{C_\mathrm{om}}
\newcommand{\Cem}{C_\mathrm{em}}
\newcommand{\bigO}{\mathcal{O}}
\newcommand{\nadd}{n_\mathrm{add}}

\newcommand{\Equbit}{\mathcal{E}_\mathrm{qubit}}
\newcommand{\gomopi}{g_\mathrm{om}/(2\pi)}
\newcommand{\gem}{g_\mathrm{em}}
\newcommand{\gemopi}{g_\mathrm{em}/(2\pi)}
\newcommand{\gemtilde}{\tilde{g}_\mathrm{em}}
\newcommand{\Phii}{\Phi_\mathrm{i}}

\newcommand{\kappao}{\kappa_{\rm o}}
\newcommand{\kappaoi}{\kappa_{\rm o,i}}
\newcommand{\kappaoe}{\kappa_{\rm o,e}}

\newcommand{\kappamue}{\kappa_{\mu\rm,e}}
\newcommand{\kappamu}{\kappa_{ \mu}}
\newcommand{\gammam}{\gamma_\mathrm{m}}

\newcommand{\etao}{\eta_{\rm o}}
\newcommand{\etaem}{\eta_\mathrm{em}}
\newcommand{\etamu}{\eta_\mu}

\usepackage{afterpage}

\makeatletter
\def\@email#1#2{%
 \endgroup
 \patchcmd{\titleblock@produce}
  {\frontmatter@RRAPformat}
  {\frontmatter@RRAPformat{\produce@RRAP{*#1\href{mailto:#2}{#2}}}\frontmatter@RRAPformat}
  {}{}
}%
\makeatother
\begin{document}

\preprint{AIP/123-QED}

\title[]{Design of a release-free piezo-optomechanical quantum transducer}
\author{Paul Burger}
    \altaffiliation[]{paulbu@chalmers.se}
\author{Joey Frey}%
\altaffiliation[]{Similarly contributed.}
\author{Johan Kolvik}
\altaffiliation[]{Similarly contributed.}
\author{David Hambraeus}
\author{Rapha\"{e}l Van Laer$^*$}
    \altaffiliation[]{raphael.van.laer@chalmers.se}%
\affiliation{ 
Department of Microtechnology and Nanoscience (MC2), Chalmers University of Technology
}%



\date{\today}

\begin{abstract}
Quantum transduction between microwave and optical photons could combine the long-range connectivity provided by optical photons with the deterministic quantum operations of superconducting microwave qubits. A promising approach to quantum microwave-optics transduction uses an intermediary mechanical mode along with piezo-optomechanical interactions. So far, such transducers have been released from their underlying substrate to confine mechanical fields -- preventing proper thermal anchoring and creating a noise-efficiency trade-off resulting from optical absorption. Here, we introduce a release-free, i.e. non-suspended, piezo-optomechanical transducer intended to circumvent this noise-efficiency trade-off. We propose and design a silicon-on-sapphire (SOS) release-free transducer with appealing piezo- and optomechanical performance. Our proposal integrates release-free lithium niobate electromechanical crystals with silicon optomechanical crystals on a sapphire substrate meant to improve thermal anchoring along with microwave and mechanical coherence. It leverages high-wavevector mechanical modes firmly guided on the chip surface. Beyond quantum science and engineering, the proposed platform and design principles are attractive for low-power acousto-optic systems in integrated photonics.
\end{abstract}

\maketitle

\section{Introduction}

Optical and microwave photons each have distinct attractive properties. Superconducting microwave qubits support near-deterministic coherent quantum operations \cite{kjaergaard_superconducting_2020}. Optical photons have vanishing thermal occupation even at room temperature and enable long-range connectivity. Similarly to classical optical interconnects \cite{miller_attojoule_2017,shekhar_roadmapping_2024,zhu_integrated_2021}, quantum microwave-optics transducers promise to connect these complementary fields -- bridging five orders of magnitude in operating frequency through optically pumped up- or down-converting three-wave-mixing interactions. If sufficiently performant, these transducers would unlock demonstrations of optically-mediated entanglement between superconducting qubits \cite{krastanov_optically_2021} and microwave-powered optical entanglement \cite{haug_heralding_2024}. This could impact quantum computing, communication, and sensing \cite{clerk_hybrid_2020,lauk_perspectives_2020}.

\begin{figure*}[ht!]
\centering\includegraphics[]{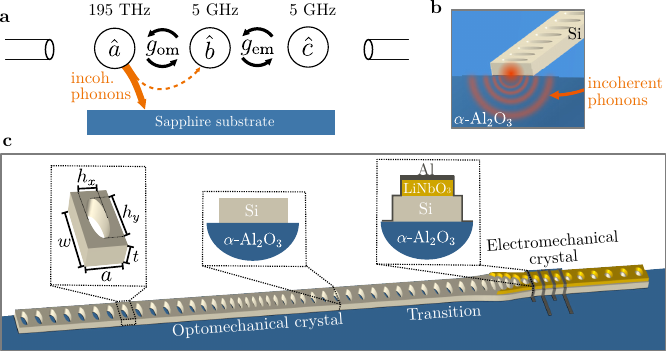}
\caption{\textbf{Proposal and design of a release-free silicon-on-sapphire (SOS) piezo-optomechanical transducer.} \textbf{a)} Principle of the transduction process. An optical mode $\hat{a}$ and a microwave mode $\hat{c}$ are coupled via an intermediary mechanical mode $\hat{b}$. The input and output lines are evanescently coupled to the optical and microwave cavity modes. Incoherent phonons created by parasitic absorption heating in the optical mode can end up in the substrate, as preferred, or in the mechanical mode -- reducing the signal-to-noise ratio of the transduction process. \textbf{b)} Cross-section of the proposed device in the optomechanical region: Incoherent heat phonons (red) can escape from the silicon optomechanical region directly into the substrate. \textbf{c)} Overview of the release-free transducer device. The insets depict the nomenclature of the unit cell parameters and the material stack at different locations in the device.}
\label{fig:overview}
\end{figure*}

Realizing these goals is a major challenge given the vast gap between microwave and optical frequencies and the strict performance requirements. It requires combining superconducting and optical technology in a cryogenic environment without reducing their performance and doing so ideally in a scalable way. To this end, a variety of platforms including direct electro-optic, bulk electro-optomechanical, and atom- or magnon-based systems are under development \cite{hease_bidirectional_2020,higginbotham_harnessing_2018,doeleman_brillouin_2023,brubaker_optomechanical_2022,han_microwave-optical_2021}. Harnessing piezo-optomechanical interactions is a promising path because of the wavelength-scale overlap between optical and mechanical fields. This results in strong three-wave-mixing interaction rates and thus low energy dissipation in the power-constrained cryogenic environments needed for superconducting qubits. These piezo-optomechanical transducers employ an intermediary mechanical mode that interacts with the microwave and optical fields, creating a coherent and bidirectional bridge between them (Fig. \ref{fig:overview}a) \cite{safavi-naeini_controlling_2019}. Typically, the device is based on an optomechanical crystal (OMC) that co-localizes optical and mechanical fields in a micron-scale silicon beam \cite{eichenfield_optomechanical_2009}. An optical pump then facilitates photon up- and down-conversion via the linearized optomechanical interaction Hamiltonian: $\hat{\mathcal{H}}_{\rm om}=\hbar \gom \sqrt{\nc}( \hat{a}^{\dagger}+ \hat{a}) (\hat{b}+\hat{b}^{\dagger})$, with $\hat{a}$ and $\hat{b}$ the photonic and phononic ladder operators and $\gom$ the vacuum optomechanical interaction rate (Fig. \ref{fig:overview}a)\cite{aspelmeyer_cavity_2014}. The interaction can be enhanced by increasing the number of pump photons $\nc$ in the cavity. This category of transducers has improved rapidly and has seen efficiencies increase from order $10^{-5}$ to the percent level\cite{mirhosseini_superconducting_2020, jiang_optically_2023, weaver_integrated_2024,zhao_quantum-enabled_2024} in a matter of years.
These state-of-the-art piezo-optomechanical devices are suspended, i.e. released from the underlying substrate. This enables excellent mechanical confinement of a mechanical breathing mode with a high optomechanical coupling rate $\gom$. However, parasitic absorption of optical pump photons results in heating of the mechanical mode \cite{meenehan_pulsed_2015}. The suspension that allows mechanical coherence also prevents the heat from dissipating rapidly \cite{ren_two-dimensional_2020}. This gives rise to a seemingly fundamental noise-efficiency trade-off: Increasing the optical pump power increases the efficiency, but also the added noise. 
As a consequence, the piezo-optomechanical transducers are often operated using short pulses to allow the mechanical mode to cool down in between successive shots \cite{meesala_non-classical_2024, jiang_optically_2023}. Improved thermalization of the optomechanical crystal would enable quantum state transduction with higher fidelity and bandwidth for both continuous-wave and pulsed protocols.

An approach to improve thermalization uses two-dimensional optomechanical crystals where the devices are patterned on a suspended sheet of silicon rather than on a one-dimensional nanobeam \cite{safavi-naeini_two-dimensional_2014}. These structures led to about an order-of-magnitude improvement in the thermal phonon population in pure optomechanics experiments \cite{ren_two-dimensional_2020,sonar_high-efficiency_2024,mayor_two-dimensional_2024}. Although these devices show impressive improvements over one-dimensional suspended structures, they require tight dimensional control and fine-tuning along with in-plane bandgaps to support modes. This can decrease device robustness and yield \cite{safavi-naeini_two-dimensional_2014} unlike in one-dimensional structures where clean modes exist even in the presence of small geometric deviations. Achieving high-fidelity and fast microwave-optics transduction limited only by the piezo-optomechanical interaction rates and the cryostat's overall cooling power -- rather than by the local parasitic heating -- benefits from revisiting the basic design principles behind piezo-optomechanical transducers.

To address these challenges, we pursue a release-free, i.e. non-suspended, and one-dimensional architecture. The proposed devices are in full contact with their underlying substrate. This is intended to allow the optically dissipated energy to flow into the substrate before it populates the mechanical mode (Fig. \ref{fig:overview}a-b). Along this research direction, we previously reported the design and room-temperature measurement of release-free silicon optomechanical crystals with silicon dioxide as substrate \cite{kolvik_clamped_2023-1}. These structures showed an attractive vacuum optomechanical interaction rate of $\gomopi = 0.5 \ \text{MHz}$ as well as good sideband-resolution and acceptable room-temperature mechanical quality, motivating us to build on this approach.

Going beyond our prior work \cite{kolvik_clamped_2023-1}, here we design an entire release-free piezo-optomechanical transducer. Current suspended transducer designs have a nearly vanishing longitudinal mechanical wavevector, which implies that the mechanical mode would quickly be lost to the substrate if not for the suspension. To overcome this challenge, we explore an unconventional operating point for opto- and electromechanical crystals harnessing high mechanical wavevectors. This places the mechanical mode well outside the continuum of substrate modes. To achieve strong interaction rates between high-wavevector phonons and optical photons, we leverage counter-propagating three-wave-mixing processes \cite{kolvik_clamped_2023-1}.  Our reliance on high mechanical wavevectors and counter-propagating interactions for quantum transduction thus ties together the rich fields of Brillouin scattering and cavity quantum optomechanics \cite{safavi-naeini_controlling_2019,van_laer_unifying_2016,eggleton_brillouin_2019,aspelmeyer_cavity_2014}. The optomechanical crystal section of the transducer uses backward Brillouin scattering as studied mainly in traveling-wave optomechanical structures such as non-suspended chalcogenide waveguides \cite{eggleton_brillouin_2019} and partially suspended silicon nanowires \cite{van_laer_interaction_2015,qiu_stimulated_2013} so far.

We propose sapphire as a crystalline substrate with the aim of improving thermal anchoring in addition to microwave and mechanical coherence compared to the amorphous silicon dioxide substrate used in Kolvik et al.\cite{kolvik_clamped_2023-1}. Sapphire conducts heat well and supports excellent superconducting microwave and mechanical coherence as seen in cutting-edge superconducting qubit and quantum acoustics experiments \cite{place_new_2021,von_lupke_parity_2022,martinis_ucsb_2014}. On top of the sapphire, the design combines thin films of lithium niobate and silicon: lithium niobate for its strong piezoelectricity \cite{arrangoiz-arriola_resolving_2019,wollack_quantum_2022} and silicon as a leading material in optomechanics for its high refractive index and optical and mechanical quality\cite{maccabe_nano-acoustic_2020,ren_two-dimensional_2020}. Lithium niobate and silicon are also widely used in integrated photonics \cite{zhu_integrated_2021,shekhar_roadmapping_2024}. In addition, the speed of sound in sapphire exceeds that in silicon by roughly 15\% (see Appendix \ref{app:SAWs}), allowing the thin films on top of it to support confined mechanical modes even without geometric softening \cite{kolvik_clamped_2023-1}. Thus, our design uses a heterogeneous material platform with the aim of satisfying the stringent performance requirements to transduce microwave photons, phonons, and optical photons at the quantum level.

The outline is as follows. After briefly introducing the method of mechanical confinement, we design SOS optomechanical crystals with performance similar to our prior work with the silicon dioxide substrate \cite{kolvik_clamped_2023-1}. Next, we design a release-free lithium niobate on SOS electromechanical crystal (EMC). We then create a partial mirror to acoustically connect and hybridize the electro- and optomechanical crystals, leading to a release-free piezo-optomechanical transducer design. The transducer's microwave operating frequency near $\omopi\approx 5 \rm \ GHz$ is well matched to state-of-the-art superconducting qubits. Finally, we analyze the resulting transducer and assess its robustness and other metrics.

\begin{figure*}[ht!]
\centering\includegraphics[]{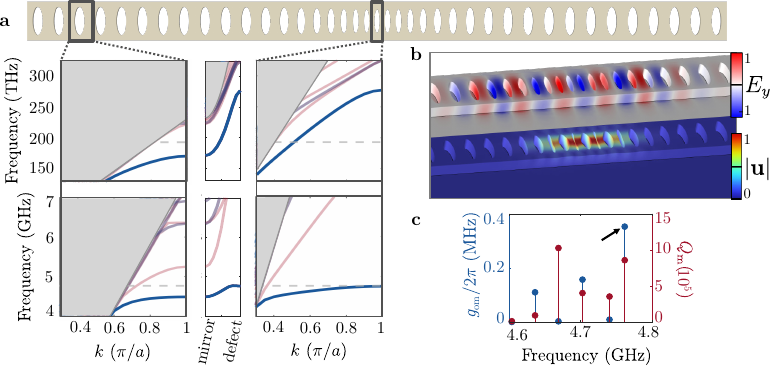}
\caption{\textbf{Release-free silicon-on-sapphire optomechanical crystal.} \textbf{a)} Optical (top) and mechanical (bottom) band structures of the nominal mirror (left) and defect cell (right) of the short OMC (sOMC). The middle plot shows the frequencies at the $X$-point as the bands are perturbed from mirror to defect. The horizontal dashed line indicates the operating frequency. \textbf{b)} Optical (top) and mechanical (bottom) mode profiles of the sOMC. \textbf{c)} Vacuum optomechanical coupling rates $\gom$ (left, blue) and radiation-limited mechanical quality factors $\Qm$ (right, red) of the sOMC modes. The arrow indicates the mode of interest.}
\label{fig:componentsOMC}
\end{figure*}

\section{Release-free mechanical confinement}
Our device consists of three subsections: the optomechanical crystal (OMC), the electomechanical crystal (EMC) and the partial mirror transition that connects and hybridizes the OMC with the EMC (Fig. \ref{fig:overview}c). Our mechanical modes are confined to the device layer in each of these sections even in the presence of the substrate, as their wavevectors exceed those of any substrate modes at the same frequency. They lose energy to acoustic modes in the substrate through rapid geometric perturbations and disorder, unlike in suspended devices -- where full bandgaps help prevent this. Through careful photonic and phononic engineering exploiting slow transitions and high wavevectors, our design aims to keep such scattering of coherent GHz phonons into the substrate under control. Our approach to mechanical confinement is similar to the confinement of optical fields through total internal reflection, in contrast to non-suspended efforts using e.g. bound states in the continuum \cite{liu_optomechanical_2022}. As in our previous release-free optomechanical crystals \cite{kolvik_clamped_2023-1}, there is also geometric softening \cite{kolvik_clamped_2023-1, sarabalis_release-free_2017,mason_ridge_1971,lagasse_higherorder_1973} near the material boundaries. This effect reduces the mechanical frequencies further below the continuum of substrate modes at fixed wavevector. These confinement principles affect only a narrow subset of the phononic modes of the system with sufficiently high wavevectors. Most phononic modes in the system have poor confinement by design. In this way, we aim to confine coherent GHz phonons while letting incoherent heat phonons flow into the substrate.

\section{Release-free optomechanical crystal}

To achieve strong interaction rates between high-wavevector phonons and optical photons, we leverage counter-propagating three-wave-mixing processes. That is, we couple high-wavevector mechanical modes near the $X$-point $\km \approx \pi/a$ to a near-infrared optical mode of half the wavevector $\ko \approx \pi/(2a)$ through phase-matched and counter-propagating three-wave-mixing, where $a$ is the unit cell period (Fig. \ref{fig:overview}c), see Kolvik et al.\cite{kolvik_clamped_2023-1} and Appendix \ref{sec:SIgom}. Thus, we use backward Brillouin scattering in a non-suspended silicon optomechanical crystal cavity \cite{eggleton_brillouin_2019}.

The desire for simultaneous confinement of mechanical and optical modes leads us to examine short unit cells with a period of $a\approx 200 \rm \ nm$ as these push the acoustic substrate continuum to higher frequencies. Like their suspended counterparts, these unit cells are patterned with ellipses -- whose band structure can be tuned geometrically; see Appendix \ref{app:freqtuning}. The unit cell with its relevant dimensions is shown in Fig. \ref{fig:overview}c. Finite-element simulations of the unit cell let us find an optical mode in the C-band near a free-space wavelength of 1550 nm and a mechanical mode around $\omopi\approx\ \rm5\ GHz$. The pinch-like\cite{eichenfield_optomechanical_2009} mechanical displacement is mostly along the beam and perpendicular to the quasi-TE optical mode. The optical field is predominantly in the dielectric. We compute a vacuum optomechanical coupling rate in the periodic unit cell of $g_{\rm om, uc}/(2\pi)=3.8\ \rm MHz$, see also Appendix \ref{sec:SIgom}.
Sapphire has a refractive index of 1.74, resulting in a minor penalty in the optical confinement with respect to a silicon dioxide substrate\cite{kolvik_clamped_2023-1}, see also Appendix \ref{app:oxidelayer}.

We create a cavity from a series of unit cells by perturbing the periods and hole sizes towards the ends of the cavity to open a bandgap for both mechanical and optical fields. Before optimizing the cavity, we ensure that the cavity supports modes with the correct wavevectors by Fourier transforming the fields \cite{kolvik_clamped_2023-1}. Next, we optimize the cavity to maximize the vacuum optomechanical coupling rate and the optical and mechanical quality factors using a Nelder-Mead algorithm, see Appendix \ref{app:optim}. We simulate radiation-limited quality factors using perfectly matched layers in the absence of disorder. They set an upper bound on our experimental quality factor expectations. The resulting cavity and band structures are shown in Fig. \ref{fig:componentsOMC}a-b.

The mechanical mode is near the band edge, while the optical mode is not. Hence, the bandgap opens for the mechanical mode at the beginning of the transition region while it takes more unit cells for the optical mode. The resulting spatial mismatch between the optical and mechanical modes shown in Fig. \ref{fig:componentsOMC}b motivates us to examine also a longer OMC (lOMC) with $N=9$ identical defect cells, see also Appendix \ref{app:cavSpectra}. This should decrease the impact of the transition region on the mode profile and reduce the spatial mismatch, bringing the optomechanical coupling rates of the full cavity closer to those expected from the coupling rate $g_{\rm om, uc}$ of the periodic unit cell. This also reduces the optical energy density, likely improving thermalization compared to the short OMC (sOMC) with $N=1$ defect cell. 
Increasing the length of the cavity by adding identical defect unit cells does not impact the maximum optomechanical cooperativity in the large-$N$ limit where $\gom$ scales only with the mechanical zero-point motion. Indeed, we then have $ \gom \propto 1/\sqrt{N}$\cite{aspelmeyer_cavity_2014,van_laer_unifying_2016,safavi-naeini_two-dimensional_2014} and $n_{\rm c,max} \propto N$, assuming thermal conductance scaling linearly with $N$ and at fixed added noise. Therefore, the maximal optomechanical cooperativity, $\Com \propto \gom^2 n_{\rm c,max}$, is independent of $N$ in this limit. 
In practice, $\gom$ benefits from longer cavities as this simplifies exploiting the strong unit-cell vacuum optomechanical coupling rate $g_{\rm om, uc}$. 
However, as we will see later, the larger mode volume for the mechanical mode decreases the hybridization in the transducer creating a trade-off between peak performance and robustness. In addition, a shorter transducer benefits from a less crowded mechanical spectrum.

\begin{figure*}[hbt!]
\centering\includegraphics[]{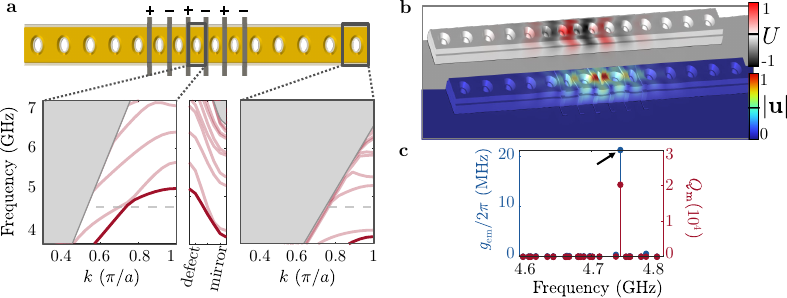}
\caption{\textbf{Release-free silicon-on-sapphire electromechanical crystal.} \textbf{a)} Mechanical band structure of defect (left) and mirror cell (right) in the electromechanical crystal (EMC). The middle plot shows the frequencies at the $X$-point as the bands are perturbed from defect to mirror. The horizontal dashed line indicates the operating frequency. \textbf{b)} Mechanical mode profile (bottom) of the EMC. The upper inset shows the electrical potential associated with the mode. The latter simulation has no electrodes in order to only show the piezoelectric potential. \textbf{c)} Electromechanical coupling rates $\gem$ (left, blue) and radiation-limited mechanical quality factors $\Qm$ (right, red) of the EMC modes. The arrow indicates the mode of interest}
\label{fig:componentsEMC}
\end{figure*}

For the optimized release-free sOMC, we simulate an optical mode with $\ooopi=195\ \text{THz}$ and a mechanical mode with $\omopi=4.8\ \text{GHz}$ with radiation-limited quality factors $\Qo=2\cdot 10^{6}$ and $\Qm=9\cdot 10^{4}$ respectively. The simulated vacuum optomechanical coupling is $\gomopi=0.35\ \text{MHz}$, with the moving-boundary term as the dominant contribution to the optomechanical interaction (Fig. \ref{fig:componentsOMC}c). For the lOMC the coupling increases to around $\gomopi=0.43\ \text{MHz}$. The optical mode volume is roughly 30\% larger than in the sOMC. The increased thermal contact area would likely result in lower heating at the same $\nc$ in the lOMC as alluded to above.

\section{Release-free electromechanical crystal}

Now we turn to designing the electromechanical crystal (EMC). We approach its design in a release-free manner as well. The alternative of combining a suspended EMC with a non-suspended OMC likely produces excessive mechanical losses at the interface between them. Furthermore, optical absorption may also occur in the EMC region, e.g. near the electrodes, which may contribute to noise in the transducer. We therefore leave the EMC on the substrate and harness similar mechanical design principles including high mechanical wavevectors as for the OMC.

Considering the EMC defect unit cell, we look for modes with frequencies and shapes similar to those in the OMC. Similar mode profiles would let us avoid a mode-converting transition, such that it may be easier to achieve a large mechanical overlap and stronger hybridization between the EMC and OMC mechanical modes (Sec. \ref{sec:EMCOMC}). Both the desire for similar mode profiles as well as the ability to control the bands to confine the mode lead us to examine unit cells similar to those in the OMC. The material stack is a $100\ \rm nm$ thin film of lithium niobate (LN) on top of the same SOS stack as in the OMC. We choose lithium niobate as a strong piezoelectric used with success in state-of-the-art quantum acoustics experiments using suspended GHz phononic crystals \cite{arrangoiz-arriola_resolving_2019,wollack_loss_2021}. The LN is slightly narrower than the silicon beam to leave sufficient margin for misalignment in fabrication. It is deliberately chosen to be thin to ease the transition from EMC to OMC (Sec. \ref{sec:transition}) and to reduce the impact of the sidewalls present in the LN etch.

We use an inter-digitated capacitor on top of the beam to actuate the piezomechanical mode (Fig. \ref{fig:componentsEMC}a). We consider aluminum electrodes here. Other materials with beneficial properties such has high kinetic inductance and low quasiparticle lifetime can also be used \cite{mckenna_cryogenic_2020,meesala_non-classical_2024}. The static capacitance of the microwave field in the LN should be small to reduce the detrimental impact of the LN's microwave loss tangent on the microwave mode's coherence, see also Appendix \ref{app:gem} \cite{zorzetti_millikelvin_2023,wollack_loss_2021,chiappina_design_2023-1}. At the same time, the LN region needs to be large enough to yield a good electromechanical interaction rate. As another precaution to improve the transition (see Sec.\ref{sec:EMCOMC}), the elliptical hole is patterned into both the LN and the silicon underneath, taking into account margins for misalignment (see Appendix \ref{sec:SIparams}).

The electromechanical coupling depends on the crystal orientation. We choose an orientation such that the piezoelectric coupling matrix has a large $e_{11}$ component, $e_{11}=-4\ \rm C/m^2$. This component connects the field in \textit{x}-direction generated by the inter-digitated capacitor to the dominant source of strain, $S_{11}$, of our mechanical mode (see also Appendix \ref{app:gem}). We examine the electric potential of the mechanical modes arising from piezoelectricity (Fig. \ref{fig:componentsEMC} b), confirming that the microwave field impinging on the electrodes properly addresses it. Thus, we find a mechanical mode with $\omega_{\rm m}/(2\pi)=4.7\ \text{GHz}$ with an electromechanical coupling rate of $g_{\text{em}}/(2\pi)=21\ \text{MHz}$, when coupled to a $C_\mu=70\ \text{fF}$, i.e. impedance of $Z_\mu=478\ \Omega$, microwave resonator or qubit (see also Appendix \ref{app:gem}). This simulated piezoelectric interaction rate is as strong as in suspended lithium niobate phononic defects \cite{arrangoiz-arriola_resolving_2019}. The static capacitance for three electrode periods is $C_{\rm IDT}= 0.5\ \rm fF $ -- of a magnitude similar to that in suspended structures \cite{chiappina_design_2023-1,arrangoiz-arriola_resolving_2019}. Using 1 mirror and 9 transition cells is sufficient to achieve a radiation-limited quality factor $Q_{\rm m}=2\cdot 10^{4}$. The band diagram is shown in Fig. \ref{fig:componentsEMC}a. The unit cell simulations do not include electrodes. Adding electrodes to the EMC cavity reduces the mechanical frequency by about $100\ \rm MHz$.

\section{Transition between opto- and electromechanical crystals}
\label{sec:transition}
To realize a piezo-optomechanical transducer, we connect the opto- and electromechanical cavities via a partial mirror region. We design this transition region to act as a phononic waveguide at $\approx5\ \rm GHz$ while simultaneously supporting a quasi-bandgap for the optical field. 
\begin{figure}[ht!]
\centering\includegraphics[]{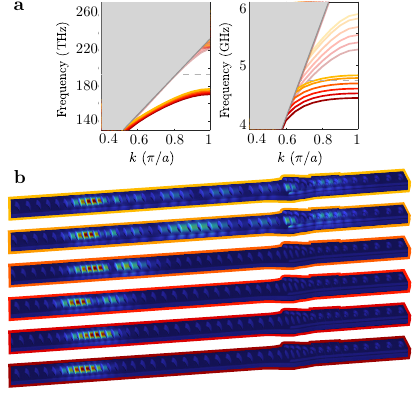}
\caption{\textbf{Band and mode evolution for the partial mirror.} \textbf{a)} Optical (left) and mechanical (right) band structures as the unit cell is transformed from mirror (dark red) to partial mirror (orange). We provide the unit cell parameters in Appendix \ref{sec:SIparams}. \textbf{b)} Full transducer mode profiles for the partial mirror transformation in a).} 
\label{fig:partialMirrorsweep}
\end{figure}
We visualize (Fig. \ref{fig:partialMirrorsweep}) how the bands of a unit cell and mode profiles in the full transducer -- to be introduced in the following section -- change as it is transformed from a full mirror to a partial mirror.
As the bands in the partial mirror get closer to the OMC mechanical frequency, the mechanical mode expands into the partial mirror direction. When the partial mirror band surpasses the OMC mechanical frequency, we find a propagating mode in the partial mirror which hybridizes the OMC and EMC modes. Before that point, the energy participation ratio in the EMC is negligible at $\ll 1\%$. This means that the existence of a mode in the partial mirror is essential for the hybridization. At the same time, the band must not be lifted so far beyond the operating frequency that the mode could leak into the continuum.
A partial mirror region is also present in the suspended devices. In that case, the phononic waveguide is terminated by an unpatterned suspended block of silicon with a piezoelectric material on top. In our non-suspended device, however, the sudden change in geometry would cause scattering losses at the LN to no-LN interface. Instead, we adiabatically guide the mechanical mode from LN into the silicon. This comes at the cost of a longer transition region.

\begin{figure*}[ht!]
\centering\includegraphics[]{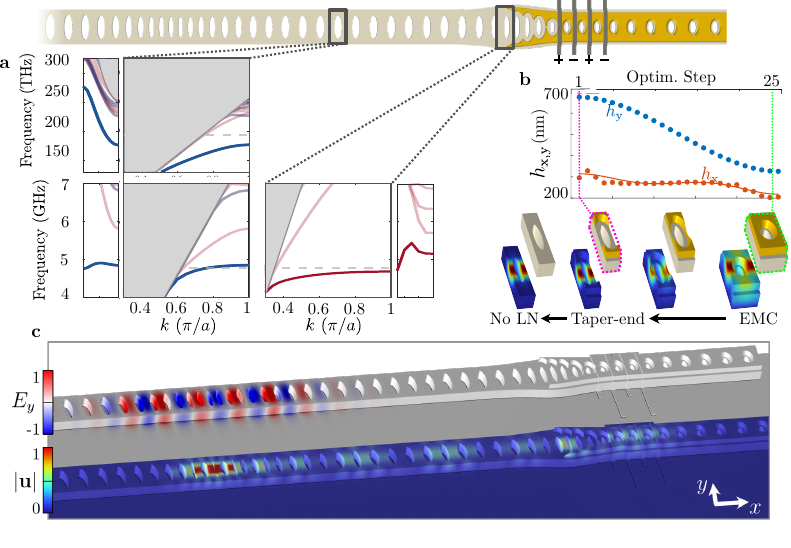}
\caption{\textbf{Release-free silicon-on-sapphire piezo-optomechanical transducer through hybridization of opto- and electromechanical crystals.} \textbf{a)} Band diagrams of the partial mirror cell (left) and the taper-end cell (right). The narrow inset figures show the $X$-point frequency as the unit cell is perturbed to the OMC defect cell and to the EMC defect cell respectively. \textbf{b)} Principle of the taper transition: starting from the EMC defect cell the LN is slowly removed to the sides which transfers the acoustic energy to the silicon. The plot shows how the LN ellipse parameters $h_x$ and $h_y$ vary along the transition. The points in the plot corresponding to the presented cells are shown with dashed lines. \textbf{c)} Optical (top) and mechanical (bottom) mode profile of the final sOMC-based transducer design.}
\label{fig:transducer}
\end{figure*}

The LN-to-Si transition region follows the following procedure: Starting from the EMC defect we gradually taper the mode from the LN into the silicon. This is done by simultaneously decreasing period $a$, increasing width $w$ and expanding the LN ellipse by increasing $h_y$ and $h_x$. We let the ellipses overlap so that the LN slowly recedes to the sides of the beam (Fig. \ref{fig:transducer}b). We do not allow for disconnected pieces of LN since they might be difficult to fabricate. 

The taper region is defined by the parameters of the EMC defect cell and the final taper cell before the LN is removed. The final taper cell, also referred to as taper-end cell, is designed such that it has low energy participation in the LN. The transition between the two defining cells is made over several unit cells where all parameters with the exception of $h_x$ are varied according to a cubic function, just as for the transition regions in OMC and EMC \cite{chan_laser_nodate}. The ellipse parameter $h_x$ is transformed according to a polynomial function (Fig. \ref{fig:transducer}a). 
This tapered version of the EMC has only two electrode periods, reducing the static capacitance to $C_{\rm IDT}= 0.3\ \rm fF $.

\section{Release-free piezo-optomechanical transducer}
\label{sec:EMCOMC}
Next, we assemble the OMC, EMC, and transition sections designed above into two release-free piezo-optomechanical transducers: a sOMC- and a lOMC-based transducer. We show the sOMC-based transducer in Fig. \ref{fig:transducer}c. We connect these individual sections through tapering regions to slowly deform unit cells at the end of one section into unit cells at the beginning of the next section. Connecting the OMC and EMC sections changes the mechanical and optical mode frequencies. In addition, it increases the mechanical mode density as a result of the increase in the total length of the cavity.

\begin{figure*}[ht!]
\centering\includegraphics[]{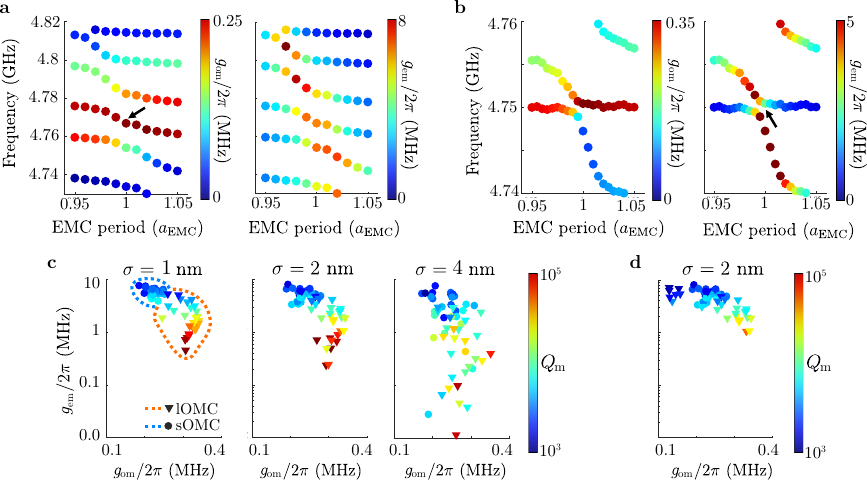}
\caption{\textbf{Opto- and electromechanical mode anti-crossings and simulated robustness of the release-free transducer versus geometric disorder.} \textbf{a)} Mechanical mode frequencies as a function of normalized EMC defect cell period for the sOMC-based transducer. The color denotes $\gom$ (left) and $\gem$ (right) respectively. Varying the period changes the frequency of the EMC mode, which exhibits an anti-crossing with the OMC mode. The black arrow denotes the mode defined in the main text. \textbf{b)} same as a) but for the lOMC-based transducer. \textbf{c)} We subject the sOMC- (circles) and lOMC-based (triangles) transducers to random disorder in their hole sizes with the given standard deviation $\sigma$ in 30 separate simulations. The mode with the highest $\gom$ is chosen as the data point. In fact, the devices usually support several modes with appreciable $\gom$ and $\gem$ (see also Fig. \ref{fig:crossing}a). \textbf{d)} same as c) but including only modes with $\gemopi \ge 1\ \rm MHz$.}
\label{fig:crossing}
\end{figure*}

We characterize the design by sweeping the EMC mechanical mode frequency for a constant OMC mechanical mode frequency. This is done by varying the EMC defect period $a$ -- revealing an anti-crossing between the EMC and OMC mechanical modes (Fig. \ref{fig:crossing}a). Having found the EMC-OMC anti-crossing, we re-optimize the entire device once again for the coupling rates and quality factors. 

The sOMC-based transducer exhibits a vacuum optomechanical coupling rate of $\gomopi=0.24\ \rm MHz$ and piezomechanical coupling rate of $\gemopi=6\ \rm MHz$ at peak hybridization. The spacing to the neighboring modes is $\approx 15\ \rm MHz$. The piezomechanical coupling rate $\gem$ drops to half of its peak value at a detuning of $\pm3\%$ in $a$, corresponding to a $\pm40\ \rm MHz$ frequency detuning of the EMC versus the OMC. In this range, the fraction of mechanical energy participating in the electromechanical region decreases from $12\%$ to $5\%$. When the participation in the lossy transition region is at its peak near the anti-crossing, the mechanical quality factor drops to $\Qm\approx3\cdot 10^3$. For the lOMC-based transducer the degree of hybridization is reduced due to the lower overlap between the evanescent fields of the two modes (Fig. \ref{fig:crossing}b). Even at the anti-crossing, the participation in the piezoelectric is only $\approx2\%$ and the mechanical quality factor increases to $\Qm\approx1\cdot 10^4$. Nonetheless, we find $\gemopi=2.4\ \rm MHz$ along with a larger $\gomopi=0.34\ \rm MHz$ for the lOMC-based transducer.

\subsection{Disorder sensitivity and robustness}
\label{sec:robustness}
To evaluate the robustness of these designs, we introduce artificial disorder. The disorder has the form of random size variations of each ellipse, i.e. its $h_x$ and $h_y$, sampled from a normal distribution centered at zero. This is meant to capture fabrication disorder resulting from e.g. the lithography and etching. Fig. \ref{fig:crossing}c shows the effect of increasing disorder on the transducer performance. Each data point corresponds to the mode with the highest $\gom$ in a given simulation. We perform these disorder simulations for both the sOMC- and lOMC-based transducers. We can distinguish following tendencies in the two transducer types: lower $\Qm$, higher $\gem$ for the sOMC-based transducer and higher $\Qm$, lower $\gem$ for the lOMC-based transducer. Despite the generally lower $\gem$, the lOMC also supports modes with high $\gem$ (Fig. \ref{fig:crossing}d).
The optical quality factor is also affected by the disorder. It decreases from $\Qo >10^6$ to an average of $\Qo\approx 2\cdot 10^5$ for $\sigma=2\ \rm nm$, see also Appendix \ref{app:disorder}. 

\subsection{Design metrics}
\label{Sec:stateOfTheArt}
In this section, our aim is to put the proposed platform and design into the context of previous designs and experimental demonstrations, to the extent possible prior to experimental work. To this end, we consider the efficiency-bandwidth product $\eta_{\rm ext}\cdot \Delta\omega$ of the transducer along with the added noise $\nadd$, where $\eta_{\rm ext}$ is the efficiency of the transducer including the coupling to external feedlines where applicable, and $\Delta\omega$ is the transducer's bandwidth. Keeping the thermal mechanical occupation and thus the added noise $\nadd$ low while maximizing the efficiency-bandwidth product $\eta_{\rm ext}\cdot \Delta\omega$ ensures that quantum states can be transduced quickly and with high single-photon signal-to-noise ratio and fidelity \cite{wang_using_2012,ren_two-dimensional_2020,han_microwave-optical_2021}.

The noise-efficiency trade-off resulting from parasitic heating implies that cutting-edge piezo-optomechanical transducers need to operate at small optomechanical cooperativity $\Com= 4\gom^2\nc/(\kappao\gammam) < 1$ to achieve low added noise $\nadd$ of e.g. order $\bigO(10^{-2})$. Here, $\nc$ is the number of intracavity pump photons and $\kappao$ the optical, $\kappamu$ the microwave, and $\gammam$ the mechanical loss rates. Under these conditions, the electromechanical cooperativity  $\Cem = 4\gem^2/(\kappamu\gammam) \gg \Com$ dominates in state-of-the-art devices such that the optomechanical interaction limits device performance. Then the efficiency-bandwidth product is linearly proportional to the maximum acceptable rate $\Phii$ at which pump photons can be dissipated:

\begin{equation}
    \begin{aligned}
    \eta_{\rm ext} \cdot \Delta\omega &=4\eta_\mu\etao\frac{\Cem\Com}{(1+\Cem+\Com)^2} \cdot \gammam(1+\Cem+\Com)\\
    &\approx 4\eta_\mu\etao\frac{\Cem\Com}{(1+\Cem)} \cdot \gammam\\
    &= 4\eta_\mu\etao\underbrace{\frac{\Cem}{(1+\Cem)}}_{\etaem} \cdot\frac{4\gom^2}{\kappao\kappaoi}  \cdot \underbrace{\kappaoi \nc}_{\Phii}\\
     &= 4 \hbar \omo \cdot \frac{\Phii}{\Equbit} \quad , \ \Equbit \equiv  \frac{\hbar \omo}{\eta_\mu\etao \etaem} \frac{\kappao\kappaoi}{4\gom^2}
    \end{aligned}
\end{equation}
with $\Phii = \kappaoi \nc$, $\etamu=\kappamue/\kappamu$, $\etao=\kappaoe/\kappao$, and $\Equbit$ the energy dissipated per transduced qubit as introduced elsewhere \cite{safavi-naeini_controlling_2019,jiang_efficient_2020}. Similarly to the energy-per-bit metric in classical optical interconnects \cite{miller_attojoule_2017,jiang_efficient_2020}, we expect the dissipated energy-per-qubit $\Equbit$ to play an important role in setting the limits to quantum data transmission in power-constrained environments like those needed for superconducting qubits. The energy dissipated per qubit $\Equbit$ is therefore inversely proportional to the efficiency-bandwidth product \textit{per dissipated pump photon}: $\eta_{\rm ext} \cdot \Delta\omega / \Phii = 4 \hbar \omo / \Equbit$. Minimizing added noise $\nadd(\Phii)$ while improving the efficiency-bandwidth product $\eta_{\rm ext} \cdot \Delta\omega$ thus benefits from reducing the energy-per-qubit $\Equbit$. As hinted at in the introduction, the locally dissipated pump photons generate a hot bath for the transducer's mechanical mode. The nature of this hot bath is under study \cite{ren_two-dimensional_2020,sonar_high-efficiency_2024,maccabe_nano-acoustic_2020}. Thus, both the efficiency-bandwidth product $\eta_{\rm ext} \cdot \Delta\omega$ and the added noise $\nadd(\Phii)$ are influenced by $\Phii$. The release-free designs presented above are intended to reduce the dependence of $\nadd(\Phii)$ on the hot bath driven by $\Phii$ such that either higher $\eta_{\rm ext} \cdot \Delta\omega$ can be achieved at fixed $\nadd$, or equivalently lower $\nadd$ can be achieved at fixed $\eta_{\rm ext} \cdot \Delta\omega$.

Learning to what extent release-free designs reduce parasitic heating requires in-depth experimental validation of our design. We estimate that our design has a similar $\Equbit$ compared to cutting-edge devices (see Appendix Tab. \ref{tab:comparison}). Here, we expect $\Cem > 10$ and therefore $\etaem > 0.9$ using the coupling rates and losses simulated here. Therefore, either $\Equbit$ needs to be reduced or the reduction in parasitic heating must be substantial for the release-free approach to surpass the suspended one in the efficiency-bandwidth product $\eta_{\rm ext} \cdot \Delta\omega$ at fixed noise level $\nadd(\Phii)$.

The area through which the optically generated heat must pass to reach the cryostat's cold bath, i.e. the thermal contact area, increases substantially by going from suspended one- and two-dimensional devices to release-free optomechanical crystals \cite{kolvik_clamped_2023-1} and to the release-free transducer designs presented here. We expect that this larger thermal contact area is associated with a larger thermal conductance and therefore a lower thermal mechanical occupation at fixed $\Phii$. This suggests that a significant reduction in parasitic heating may be accessible in release-free devices. The complex thermal physics \cite{ren_two-dimensional_2020,sonar_high-efficiency_2024,maccabe_nano-acoustic_2020} associated with parasitic heating has several key unknowns, such as millikelvin material and interface properties, requiring careful experimental investigation.

The lOMC-based transducer with a longer optomechanical section is likely to have a larger thermal contact area and thus improved thermalization compared to the sOMC-based transducer. It also has larger $\gom$ and reduced simulated radiation-limited mechanical losses. It pays for these advantages over the sOMC-based transducer with a reduction in its mechanical energy in the EMC region, leading to a decrease in hybridization and $\gem$. Given that the optomechanical three-wave-mixing is the crucial limiting factor in recent piezo-optomechanical transducer implementations, this is likely a good trade.

The presented release-free transducer may be further improved through future work focusing on the EMC and transition region, as the latter limits the simulated mechanical quality factors. The design may eventually benefit from the use of even thinner piezoelectrics to simplify the EMC and potentially circumvent acoustic tapering.

In the above, we illustrate the case of a continuous-wave efficiency-bandwidth product. Alternative pulsed schemes leverage the time delay of the heating from an optical pump pulse to operate the transducer device at lower noise levels before the heating kicks in. In these schemes, the bandwidth $\Delta\omega$ is set by the thermalization time of the mechanical mode instead of the total mechanical linewidth $\gammam(1+\Cem+\Com)$. Apart from that, the above discussion and expectations largely carry over to pulsed protocols to achieve optically-mediated entanglement of microwave qubits as well as microwave-powered optical entanglement \cite{krastanov_optically_2021,haug_heralding_2024,sonar_high-efficiency_2024,mckenna_cryogenic_2020}.

\section{Outlook and conclusion}
We propose and design a release-free piezo-optomechanical transducer, introducing silicon-on-sapphire as a platform for non-suspended electro- and optomechanical crystals. Our release-free approach may help address the parasitic heating limiting state-of-the-art piezo-optomechanical devices. Using an unconventional operating point with high mechanical wavevectors allows us to confine the mechanical mode in the device while still achieving attractive simulated electro- and optomechanical performance. Our design is one-dimensional; it supports modes even in the presence of sizeable geometric disorder. We study two versions of the transducer: a longer version with lower simulated acoustic scattering losses and larger substrate contact area and a shorter version with improved hybridization between the opto- and electromechanical sections. The devices may significantly weaken the trade-off between the efficiency and added noise of these quantum transducers. Future work could revisit the optomechanical crystal where improvements in the overlap of the optical and mechanical fields appear to be in reach. In addition, the design could be combined with piezo-free biased-silicon approaches \cite{zhao_quantum-enabled_2024}. The design principles could also be transferred to a variety of other platforms such as lithium niobate on sapphire \cite{mckenna_cryogenic_2020}. Unique to the release-free platform, the electro- and optomechanical sections might also be connected acoustically through the substrate in future design efforts. Key empirical questions include to what extent the thermalization improves with increased contact area to the substrate and whether the strong simulated piezomechanical coupling rate holds up in experiment. We effectively design an ultra-low-power and resonant non-suspended acousto-optic modulator compatible with transducing either classical or quantum signals. Such modulators may be useful beyond quantum science in microwave signal processing \cite{marpaung_integrated_2019,zhou_electrically_2024} and integrated photonics at large \cite{zhu_integrated_2021,shekhar_roadmapping_2024,sarabalis_optomechanical_2018,zhang_integrated-waveguide-based_2024}. 


\paragraph*{Acknowledgements.} We thank Trond Haug, Simone Gasparinetti, Dag Winkler, Hannes Pfeifer, Per Delsing, and Amir Safavi-Naeini for helpful discussions. We acknowledge support from the Knut and Alice Wallenberg foundation through the Wallenberg Centre for Quantum Technology (WACQT), from the European Research Council via Starting Grant 948265, and from the Swedish Foundation for Strategic Research via grant FFL21-0039.

\paragraph*{Contributions.} P.B. and R.V.L wrote the manuscript with input from all authors. P.B. led the numerical and conceptual development of the design with support from J.F., J.K., and R.V.L. P.B. and R.V.L. conceived of the approach with support from J.K. and J.F. J.F. and J.K. contributed similarly. D.H. provided conceptual and numerical input at a later stage. R.V.L. conceived of and supervised the project. 

\paragraph*{Data availability.} Data supporting the findings of this study are available from the corresponding authors upon reasonable request.

\paragraph*{\textbf{Bibliography}}
\bibliography{PB_ZoteroLib}

\begin{thebibliography}{70}%
\makeatletter
\providecommand \@ifxundefined [1]{%
 \@ifx{#1\undefined}
}%
\providecommand \@ifnum [1]{%
 \ifnum #1\expandafter \@firstoftwo
 \else \expandafter \@secondoftwo
 \fi
}%
\providecommand \@ifx [1]{%
 \ifx #1\expandafter \@firstoftwo
 \else \expandafter \@secondoftwo
 \fi
}%
\providecommand \natexlab [1]{#1}%
\providecommand \enquote  [1]{``#1''}%
\providecommand \bibnamefont  [1]{#1}%
\providecommand \bibfnamefont [1]{#1}%
\providecommand \citenamefont [1]{#1}%
\providecommand \href@noop [0]{\@secondoftwo}%
\providecommand \href [0]{\begingroup \@sanitize@url \@href}%
\providecommand \@href[1]{\@@startlink{#1}\@@href}%
\providecommand \@@href[1]{\endgroup#1\@@endlink}%
\providecommand \@sanitize@url [0]{\catcode `\\12\catcode `\$12\catcode `\&12\catcode `\#12\catcode `\^12\catcode `\_12\catcode `\%12\relax}%
\providecommand \@@startlink[1]{}%
\providecommand \@@endlink[0]{}%
\providecommand \url  [0]{\begingroup\@sanitize@url \@url }%
\providecommand \@url [1]{\endgroup\@href {#1}{\urlprefix }}%
\providecommand \urlprefix  [0]{URL }%
\providecommand \Eprint [0]{\href }%
\providecommand \doibase [0]{http://dx.doi.org/}%
\providecommand \selectlanguage [0]{\@gobble}%
\providecommand \bibinfo  [0]{\@secondoftwo}%
\providecommand \bibfield  [0]{\@secondoftwo}%
\providecommand \translation [1]{[#1]}%
\providecommand \BibitemOpen [0]{}%
\providecommand \bibitemStop [0]{}%
\providecommand \bibitemNoStop [0]{.\EOS\space}%
\providecommand \EOS [0]{\spacefactor3000\relax}%
\providecommand \BibitemShut  [1]{\csname bibitem#1\endcsname}%
\let\auto@bib@innerbib\@empty
\bibitem [{\citenamefont {Kjaergaard}\ \emph {et~al.}(2020)\citenamefont {Kjaergaard}, \citenamefont {Schwartz}, \citenamefont {Braum{\"u}ller}, \citenamefont {Krantz}, \citenamefont {Wang}, \citenamefont {Gustavsson},\ and\ \citenamefont {Oliver}}]{kjaergaard_superconducting_2020}%
  \BibitemOpen
  \bibfield  {author} {\bibinfo {author} {\bibfnamefont {M.}~\bibnamefont {Kjaergaard}}, \bibinfo {author} {\bibfnamefont {M.~E.}\ \bibnamefont {Schwartz}}, \bibinfo {author} {\bibfnamefont {J.}~\bibnamefont {Braum{\"u}ller}}, \bibinfo {author} {\bibfnamefont {P.}~\bibnamefont {Krantz}}, \bibinfo {author} {\bibfnamefont {J.~I.-J.}\ \bibnamefont {Wang}}, \bibinfo {author} {\bibfnamefont {S.}~\bibnamefont {Gustavsson}}, \ and\ \bibinfo {author} {\bibfnamefont {W.~D.}\ \bibnamefont {Oliver}},\ }\bibfield  {title} {\enquote {\bibinfo {title} {Superconducting {{Qubits}}: {{Current State}} of {{Play}}},}\ }\href {\doibase 10.1146/annurev-conmatphys-031119-050605} {\bibfield  {journal} {\bibinfo  {journal} {Annual Review of Condensed Matter Physics}\ }\textbf {\bibinfo {volume} {11}},\ \bibinfo {pages} {369--395} (\bibinfo {year} {2020})}\BibitemShut {NoStop}%
\bibitem [{\citenamefont {Miller}(2017)}]{miller_attojoule_2017}%
  \BibitemOpen
  \bibfield  {author} {\bibinfo {author} {\bibfnamefont {D.~A.~B.}\ \bibnamefont {Miller}},\ }\bibfield  {title} {\enquote {\bibinfo {title} {Attojoule {{Optoelectronics}} for {{Low-Energy Information Processing}} and {{Communications}}},}\ }\href {\doibase 10.1109/JLT.2017.2647779} {\bibfield  {journal} {\bibinfo  {journal} {Journal of Lightwave Technology}\ }\textbf {\bibinfo {volume} {35}},\ \bibinfo {pages} {346--396} (\bibinfo {year} {2017})}\BibitemShut {NoStop}%
\bibitem [{\citenamefont {Shekhar}\ \emph {et~al.}(2024)\citenamefont {Shekhar}, \citenamefont {Bogaerts}, \citenamefont {Chrostowski}, \citenamefont {Bowers}, \citenamefont {Hochberg}, \citenamefont {Soref},\ and\ \citenamefont {Shastri}}]{shekhar_roadmapping_2024}%
  \BibitemOpen
  \bibfield  {author} {\bibinfo {author} {\bibfnamefont {S.}~\bibnamefont {Shekhar}}, \bibinfo {author} {\bibfnamefont {W.}~\bibnamefont {Bogaerts}}, \bibinfo {author} {\bibfnamefont {L.}~\bibnamefont {Chrostowski}}, \bibinfo {author} {\bibfnamefont {J.~E.}\ \bibnamefont {Bowers}}, \bibinfo {author} {\bibfnamefont {M.}~\bibnamefont {Hochberg}}, \bibinfo {author} {\bibfnamefont {R.}~\bibnamefont {Soref}}, \ and\ \bibinfo {author} {\bibfnamefont {B.~J.}\ \bibnamefont {Shastri}},\ }\bibfield  {title} {\enquote {\bibinfo {title} {Roadmapping the next generation of silicon photonics},}\ }\href {\doibase 10.1038/s41467-024-44750-0} {\bibfield  {journal} {\bibinfo  {journal} {Nature Communications}\ }\textbf {\bibinfo {volume} {15}},\ \bibinfo {pages} {751} (\bibinfo {year} {2024})}\BibitemShut {NoStop}%
\bibitem [{\citenamefont {Zhu}\ \emph {et~al.}(2021)\citenamefont {Zhu}, \citenamefont {Shao}, \citenamefont {Yu}, \citenamefont {Cheng}, \citenamefont {Desiatov}, \citenamefont {Xin}, \citenamefont {Hu}, \citenamefont {Holzgrafe}, \citenamefont {Ghosh}, \citenamefont {{Shams-Ansari}}, \citenamefont {Puma}, \citenamefont {Sinclair}, \citenamefont {Reimer}, \citenamefont {Zhang},\ and\ \citenamefont {Lon{\v c}ar}}]{zhu_integrated_2021}%
  \BibitemOpen
  \bibfield  {author} {\bibinfo {author} {\bibfnamefont {D.}~\bibnamefont {Zhu}}, \bibinfo {author} {\bibfnamefont {L.}~\bibnamefont {Shao}}, \bibinfo {author} {\bibfnamefont {M.}~\bibnamefont {Yu}}, \bibinfo {author} {\bibfnamefont {R.}~\bibnamefont {Cheng}}, \bibinfo {author} {\bibfnamefont {B.}~\bibnamefont {Desiatov}}, \bibinfo {author} {\bibfnamefont {C.~J.}\ \bibnamefont {Xin}}, \bibinfo {author} {\bibfnamefont {Y.}~\bibnamefont {Hu}}, \bibinfo {author} {\bibfnamefont {J.}~\bibnamefont {Holzgrafe}}, \bibinfo {author} {\bibfnamefont {S.}~\bibnamefont {Ghosh}}, \bibinfo {author} {\bibfnamefont {A.}~\bibnamefont {{Shams-Ansari}}}, \bibinfo {author} {\bibfnamefont {E.}~\bibnamefont {Puma}}, \bibinfo {author} {\bibfnamefont {N.}~\bibnamefont {Sinclair}}, \bibinfo {author} {\bibfnamefont {C.}~\bibnamefont {Reimer}}, \bibinfo {author} {\bibfnamefont {M.}~\bibnamefont {Zhang}}, \ and\ \bibinfo {author} {\bibfnamefont {M.}~\bibnamefont {Lon{\v c}ar}},\ }\bibfield  {title} {\enquote {\bibinfo {title} {Integrated
  photonics on thin-film lithium niobate},}\ }\href {\doibase 10.1364/AOP.411024} {\bibfield  {journal} {\bibinfo  {journal} {Advances in Optics and Photonics}\ }\textbf {\bibinfo {volume} {13}},\ \bibinfo {pages} {242} (\bibinfo {year} {2021})}\BibitemShut {NoStop}%
\bibitem [{\citenamefont {Krastanov}\ \emph {et~al.}(2021)\citenamefont {Krastanov}, \citenamefont {Raniwala}, \citenamefont {Holzgrafe}, \citenamefont {Jacobs}, \citenamefont {Lon{\v c}ar}, \citenamefont {Reagor},\ and\ \citenamefont {Englund}}]{krastanov_optically_2021}%
  \BibitemOpen
  \bibfield  {author} {\bibinfo {author} {\bibfnamefont {S.}~\bibnamefont {Krastanov}}, \bibinfo {author} {\bibfnamefont {H.}~\bibnamefont {Raniwala}}, \bibinfo {author} {\bibfnamefont {J.}~\bibnamefont {Holzgrafe}}, \bibinfo {author} {\bibfnamefont {K.}~\bibnamefont {Jacobs}}, \bibinfo {author} {\bibfnamefont {M.}~\bibnamefont {Lon{\v c}ar}}, \bibinfo {author} {\bibfnamefont {M.~J.}\ \bibnamefont {Reagor}}, \ and\ \bibinfo {author} {\bibfnamefont {D.~R.}\ \bibnamefont {Englund}},\ }\bibfield  {title} {\enquote {\bibinfo {title} {Optically {{Heralded Entanglement}} of {{Superconducting Systems}} in {{Quantum Networks}}},}\ }\href {\doibase 10.1103/PhysRevLett.127.040503} {\bibfield  {journal} {\bibinfo  {journal} {Physical Review Letters}\ }\textbf {\bibinfo {volume} {127}},\ \bibinfo {pages} {040503} (\bibinfo {year} {2021})}\BibitemShut {NoStop}%
\bibitem [{\citenamefont {Haug}, \citenamefont {Kockum},\ and\ \citenamefont {Van~Laer}(2024)}]{haug_heralding_2024}%
  \BibitemOpen
  \bibfield  {author} {\bibinfo {author} {\bibfnamefont {T.~H.}\ \bibnamefont {Haug}}, \bibinfo {author} {\bibfnamefont {A.~F.}\ \bibnamefont {Kockum}}, \ and\ \bibinfo {author} {\bibfnamefont {R.}~\bibnamefont {Van~Laer}},\ }\href@noop {} {\enquote {\bibinfo {title} {Heralding entangled optical photons from a microwave quantum processor},}\ } (\bibinfo {year} {2024}),\ \Eprint {http://arxiv.org/abs/2308.14173} {arXiv:2308.14173 [quant-ph]} \BibitemShut {NoStop}%
\bibitem [{\citenamefont {Clerk}\ \emph {et~al.}(2020)\citenamefont {Clerk}, \citenamefont {Lehnert}, \citenamefont {Bertet}, \citenamefont {Petta},\ and\ \citenamefont {Nakamura}}]{clerk_hybrid_2020}%
  \BibitemOpen
  \bibfield  {author} {\bibinfo {author} {\bibfnamefont {A.~A.}\ \bibnamefont {Clerk}}, \bibinfo {author} {\bibfnamefont {K.~W.}\ \bibnamefont {Lehnert}}, \bibinfo {author} {\bibfnamefont {P.}~\bibnamefont {Bertet}}, \bibinfo {author} {\bibfnamefont {J.~R.}\ \bibnamefont {Petta}}, \ and\ \bibinfo {author} {\bibfnamefont {Y.}~\bibnamefont {Nakamura}},\ }\bibfield  {title} {\enquote {\bibinfo {title} {Hybrid quantum systems with circuit quantum electrodynamics},}\ }\href {\doibase 10.1038/s41567-020-0797-9} {\bibfield  {journal} {\bibinfo  {journal} {Nature Physics}\ }\textbf {\bibinfo {volume} {16}},\ \bibinfo {pages} {257--267} (\bibinfo {year} {2020})}\BibitemShut {NoStop}%
\bibitem [{\citenamefont {Lauk}\ \emph {et~al.}(2020)\citenamefont {Lauk}, \citenamefont {Sinclair}, \citenamefont {Barzanjeh}, \citenamefont {Covey}, \citenamefont {Saffman}, \citenamefont {Spiropulu},\ and\ \citenamefont {Simon}}]{lauk_perspectives_2020}%
  \BibitemOpen
  \bibfield  {author} {\bibinfo {author} {\bibfnamefont {N.}~\bibnamefont {Lauk}}, \bibinfo {author} {\bibfnamefont {N.}~\bibnamefont {Sinclair}}, \bibinfo {author} {\bibfnamefont {S.}~\bibnamefont {Barzanjeh}}, \bibinfo {author} {\bibfnamefont {J.~P.}\ \bibnamefont {Covey}}, \bibinfo {author} {\bibfnamefont {M.}~\bibnamefont {Saffman}}, \bibinfo {author} {\bibfnamefont {M.}~\bibnamefont {Spiropulu}}, \ and\ \bibinfo {author} {\bibfnamefont {C.}~\bibnamefont {Simon}},\ }\bibfield  {title} {\enquote {\bibinfo {title} {Perspectives on quantum transduction},}\ }\href {\doibase 10.1088/2058-9565/ab788a} {\bibfield  {journal} {\bibinfo  {journal} {Quantum Science and Technology}\ }\textbf {\bibinfo {volume} {5}},\ \bibinfo {pages} {020501} (\bibinfo {year} {2020})}\BibitemShut {NoStop}%
\bibitem [{\citenamefont {Hease}\ \emph {et~al.}(2020)\citenamefont {Hease}, \citenamefont {Rueda}, \citenamefont {Sahu}, \citenamefont {Wulf}, \citenamefont {Arnold}, \citenamefont {Schwefel},\ and\ \citenamefont {Fink}}]{hease_bidirectional_2020}%
  \BibitemOpen
  \bibfield  {author} {\bibinfo {author} {\bibfnamefont {W.}~\bibnamefont {Hease}}, \bibinfo {author} {\bibfnamefont {A.}~\bibnamefont {Rueda}}, \bibinfo {author} {\bibfnamefont {R.}~\bibnamefont {Sahu}}, \bibinfo {author} {\bibfnamefont {M.}~\bibnamefont {Wulf}}, \bibinfo {author} {\bibfnamefont {G.}~\bibnamefont {Arnold}}, \bibinfo {author} {\bibfnamefont {H.~G.}\ \bibnamefont {Schwefel}}, \ and\ \bibinfo {author} {\bibfnamefont {J.~M.}\ \bibnamefont {Fink}},\ }\bibfield  {title} {\enquote {\bibinfo {title} {Bidirectional {{Electro-Optic Wavelength Conversion}} in the {{Quantum Ground State}}},}\ }\href {\doibase 10.1103/PRXQuantum.1.020315} {\bibfield  {journal} {\bibinfo  {journal} {PRX Quantum}\ }\textbf {\bibinfo {volume} {1}},\ \bibinfo {pages} {020315} (\bibinfo {year} {2020})}\BibitemShut {NoStop}%
\bibitem [{\citenamefont {Higginbotham}\ \emph {et~al.}(2018)\citenamefont {Higginbotham}, \citenamefont {Burns}, \citenamefont {Urmey}, \citenamefont {Peterson}, \citenamefont {Kampel}, \citenamefont {Brubaker}, \citenamefont {Smith}, \citenamefont {Lehnert},\ and\ \citenamefont {Regal}}]{higginbotham_harnessing_2018}%
  \BibitemOpen
  \bibfield  {author} {\bibinfo {author} {\bibfnamefont {A.~P.}\ \bibnamefont {Higginbotham}}, \bibinfo {author} {\bibfnamefont {P.~S.}\ \bibnamefont {Burns}}, \bibinfo {author} {\bibfnamefont {M.~D.}\ \bibnamefont {Urmey}}, \bibinfo {author} {\bibfnamefont {R.~W.}\ \bibnamefont {Peterson}}, \bibinfo {author} {\bibfnamefont {N.~S.}\ \bibnamefont {Kampel}}, \bibinfo {author} {\bibfnamefont {B.~M.}\ \bibnamefont {Brubaker}}, \bibinfo {author} {\bibfnamefont {G.}~\bibnamefont {Smith}}, \bibinfo {author} {\bibfnamefont {K.~W.}\ \bibnamefont {Lehnert}}, \ and\ \bibinfo {author} {\bibfnamefont {C.~A.}\ \bibnamefont {Regal}},\ }\bibfield  {title} {\enquote {\bibinfo {title} {Harnessing electro-optic correlations in an efficient mechanical converter},}\ }\href {\doibase 10.1038/s41567-018-0210-0} {\bibfield  {journal} {\bibinfo  {journal} {Nature Physics}\ }\textbf {\bibinfo {volume} {14}},\ \bibinfo {pages} {1038--1042} (\bibinfo {year} {2018})}\BibitemShut {NoStop}%
\bibitem [{\citenamefont {Doeleman}\ \emph {et~al.}(2023)\citenamefont {Doeleman}, \citenamefont {Schatteburg}, \citenamefont {Benevides}, \citenamefont {Vollenweider}, \citenamefont {Macri},\ and\ \citenamefont {Chu}}]{doeleman_brillouin_2023}%
  \BibitemOpen
  \bibfield  {author} {\bibinfo {author} {\bibfnamefont {H.~M.}\ \bibnamefont {Doeleman}}, \bibinfo {author} {\bibfnamefont {T.}~\bibnamefont {Schatteburg}}, \bibinfo {author} {\bibfnamefont {R.}~\bibnamefont {Benevides}}, \bibinfo {author} {\bibfnamefont {S.}~\bibnamefont {Vollenweider}}, \bibinfo {author} {\bibfnamefont {D.}~\bibnamefont {Macri}}, \ and\ \bibinfo {author} {\bibfnamefont {Y.}~\bibnamefont {Chu}},\ }\bibfield  {title} {\enquote {\bibinfo {title} {Brillouin optomechanics in the quantum ground state},}\ }\href {\doibase 10.1103/PhysRevResearch.5.043140} {\bibfield  {journal} {\bibinfo  {journal} {Physical Review Research}\ }\textbf {\bibinfo {volume} {5}},\ \bibinfo {pages} {043140} (\bibinfo {year} {2023})}\BibitemShut {NoStop}%
\bibitem [{\citenamefont {Brubaker}\ \emph {et~al.}(2022)\citenamefont {Brubaker}, \citenamefont {Kindem}, \citenamefont {Urmey}, \citenamefont {Mittal}, \citenamefont {Delaney}, \citenamefont {Burns}, \citenamefont {Vissers}, \citenamefont {Lehnert},\ and\ \citenamefont {Regal}}]{brubaker_optomechanical_2022}%
  \BibitemOpen
  \bibfield  {author} {\bibinfo {author} {\bibfnamefont {B.~M.}\ \bibnamefont {Brubaker}}, \bibinfo {author} {\bibfnamefont {J.~M.}\ \bibnamefont {Kindem}}, \bibinfo {author} {\bibfnamefont {M.~D.}\ \bibnamefont {Urmey}}, \bibinfo {author} {\bibfnamefont {S.}~\bibnamefont {Mittal}}, \bibinfo {author} {\bibfnamefont {R.~D.}\ \bibnamefont {Delaney}}, \bibinfo {author} {\bibfnamefont {P.~S.}\ \bibnamefont {Burns}}, \bibinfo {author} {\bibfnamefont {M.~R.}\ \bibnamefont {Vissers}}, \bibinfo {author} {\bibfnamefont {K.~W.}\ \bibnamefont {Lehnert}}, \ and\ \bibinfo {author} {\bibfnamefont {C.~A.}\ \bibnamefont {Regal}},\ }\bibfield  {title} {\enquote {\bibinfo {title} {Optomechanical {{Ground-State Cooling}} in a {{Continuous}} and {{Efficient Electro-Optic Transducer}}},}\ }\href {\doibase 10.1103/PhysRevX.12.021062} {\bibfield  {journal} {\bibinfo  {journal} {Physical Review X}\ }\textbf {\bibinfo {volume} {12}},\ \bibinfo {pages} {021062} (\bibinfo {year} {2022})}\BibitemShut {NoStop}%
\bibitem [{\citenamefont {Han}\ \emph {et~al.}(2021)\citenamefont {Han}, \citenamefont {Fu}, \citenamefont {Zou}, \citenamefont {Jiang},\ and\ \citenamefont {Tang}}]{han_microwave-optical_2021}%
  \BibitemOpen
  \bibfield  {author} {\bibinfo {author} {\bibfnamefont {X.}~\bibnamefont {Han}}, \bibinfo {author} {\bibfnamefont {W.}~\bibnamefont {Fu}}, \bibinfo {author} {\bibfnamefont {C.-L.}\ \bibnamefont {Zou}}, \bibinfo {author} {\bibfnamefont {L.}~\bibnamefont {Jiang}}, \ and\ \bibinfo {author} {\bibfnamefont {H.~X.}\ \bibnamefont {Tang}},\ }\bibfield  {title} {\enquote {\bibinfo {title} {Microwave-optical quantum frequency conversion},}\ }\href {\doibase 10.1364/OPTICA.425414} {\bibfield  {journal} {\bibinfo  {journal} {Optica}\ }\textbf {\bibinfo {volume} {8}},\ \bibinfo {pages} {1050} (\bibinfo {year} {2021})}\BibitemShut {NoStop}%
\bibitem [{\citenamefont {{Safavi-Naeini}}\ \emph {et~al.}(2019)\citenamefont {{Safavi-Naeini}}, \citenamefont {Van~Thourhout}, \citenamefont {Baets},\ and\ \citenamefont {Van~Laer}}]{safavi-naeini_controlling_2019}%
  \BibitemOpen
  \bibfield  {author} {\bibinfo {author} {\bibfnamefont {A.~H.}\ \bibnamefont {{Safavi-Naeini}}}, \bibinfo {author} {\bibfnamefont {D.}~\bibnamefont {Van~Thourhout}}, \bibinfo {author} {\bibfnamefont {R.}~\bibnamefont {Baets}}, \ and\ \bibinfo {author} {\bibfnamefont {R.}~\bibnamefont {Van~Laer}},\ }\bibfield  {title} {\enquote {\bibinfo {title} {Controlling phonons and photons at the wavelength scale: Integrated photonics meets integrated phononics},}\ }\href {\doibase 10.1364/OPTICA.6.000213} {\bibfield  {journal} {\bibinfo  {journal} {Optica}\ }\textbf {\bibinfo {volume} {6}},\ \bibinfo {pages} {213} (\bibinfo {year} {2019})}\BibitemShut {NoStop}%
\bibitem [{\citenamefont {Eichenfield}\ \emph {et~al.}(2009)\citenamefont {Eichenfield}, \citenamefont {Chan}, \citenamefont {Camacho}, \citenamefont {Vahala},\ and\ \citenamefont {Painter}}]{eichenfield_optomechanical_2009}%
  \BibitemOpen
  \bibfield  {author} {\bibinfo {author} {\bibfnamefont {M.}~\bibnamefont {Eichenfield}}, \bibinfo {author} {\bibfnamefont {J.}~\bibnamefont {Chan}}, \bibinfo {author} {\bibfnamefont {R.~M.}\ \bibnamefont {Camacho}}, \bibinfo {author} {\bibfnamefont {K.~J.}\ \bibnamefont {Vahala}}, \ and\ \bibinfo {author} {\bibfnamefont {O.}~\bibnamefont {Painter}},\ }\bibfield  {title} {\enquote {\bibinfo {title} {Optomechanical crystals},}\ }\href {\doibase 10.1038/nature08524} {\bibfield  {journal} {\bibinfo  {journal} {Nature}\ }\textbf {\bibinfo {volume} {462}},\ \bibinfo {pages} {78--82} (\bibinfo {year} {2009})}\BibitemShut {NoStop}%
\bibitem [{\citenamefont {Aspelmeyer}, \citenamefont {Kippenberg},\ and\ \citenamefont {Marquardt}(2014)}]{aspelmeyer_cavity_2014}%
  \BibitemOpen
  \bibfield  {author} {\bibinfo {author} {\bibfnamefont {M.}~\bibnamefont {Aspelmeyer}}, \bibinfo {author} {\bibfnamefont {T.~J.}\ \bibnamefont {Kippenberg}}, \ and\ \bibinfo {author} {\bibfnamefont {F.}~\bibnamefont {Marquardt}},\ }\bibfield  {title} {\enquote {\bibinfo {title} {Cavity optomechanics},}\ }\href {\doibase 10.1103/RevModPhys.86.1391} {\bibfield  {journal} {\bibinfo  {journal} {Reviews of Modern Physics}\ }\textbf {\bibinfo {volume} {86}},\ \bibinfo {pages} {1391--1452} (\bibinfo {year} {2014})}\BibitemShut {NoStop}%
\bibitem [{\citenamefont {Mirhosseini}\ \emph {et~al.}(2020)\citenamefont {Mirhosseini}, \citenamefont {Sipahigil}, \citenamefont {Kalaee},\ and\ \citenamefont {Painter}}]{mirhosseini_superconducting_2020}%
  \BibitemOpen
  \bibfield  {author} {\bibinfo {author} {\bibfnamefont {M.}~\bibnamefont {Mirhosseini}}, \bibinfo {author} {\bibfnamefont {A.}~\bibnamefont {Sipahigil}}, \bibinfo {author} {\bibfnamefont {M.}~\bibnamefont {Kalaee}}, \ and\ \bibinfo {author} {\bibfnamefont {O.}~\bibnamefont {Painter}},\ }\bibfield  {title} {\enquote {\bibinfo {title} {Superconducting qubit to optical photon transduction},}\ }\href {\doibase 10.1038/s41586-020-3038-6} {\bibfield  {journal} {\bibinfo  {journal} {Nature}\ }\textbf {\bibinfo {volume} {588}},\ \bibinfo {pages} {599--603} (\bibinfo {year} {2020})}\BibitemShut {NoStop}%
\bibitem [{\citenamefont {Jiang}\ \emph {et~al.}(2023)\citenamefont {Jiang}, \citenamefont {Mayor}, \citenamefont {Malik}, \citenamefont {Van~Laer}, \citenamefont {McKenna}, \citenamefont {Patel}, \citenamefont {Witmer},\ and\ \citenamefont {{Safavi-Naeini}}}]{jiang_optically_2023}%
  \BibitemOpen
  \bibfield  {author} {\bibinfo {author} {\bibfnamefont {W.}~\bibnamefont {Jiang}}, \bibinfo {author} {\bibfnamefont {F.~M.}\ \bibnamefont {Mayor}}, \bibinfo {author} {\bibfnamefont {S.}~\bibnamefont {Malik}}, \bibinfo {author} {\bibfnamefont {R.}~\bibnamefont {Van~Laer}}, \bibinfo {author} {\bibfnamefont {T.~P.}\ \bibnamefont {McKenna}}, \bibinfo {author} {\bibfnamefont {R.~N.}\ \bibnamefont {Patel}}, \bibinfo {author} {\bibfnamefont {J.~D.}\ \bibnamefont {Witmer}}, \ and\ \bibinfo {author} {\bibfnamefont {A.~H.}\ \bibnamefont {{Safavi-Naeini}}},\ }\bibfield  {title} {\enquote {\bibinfo {title} {Optically heralded microwave photon addition},}\ }\href {\doibase 10.1038/s41567-023-02129-w} {\bibfield  {journal} {\bibinfo  {journal} {Nature Physics}\ }\textbf {\bibinfo {volume} {19}},\ \bibinfo {pages} {1423--1428} (\bibinfo {year} {2023})}\BibitemShut {NoStop}%
\bibitem [{\citenamefont {Weaver}\ \emph {et~al.}(2024)\citenamefont {Weaver}, \citenamefont {Duivestein}, \citenamefont {Bernasconi}, \citenamefont {Scharmer}, \citenamefont {Lemang}, \citenamefont {van Thiel}, \citenamefont {Hijazi}, \citenamefont {Hensen}, \citenamefont {Gr{\"o}blacher},\ and\ \citenamefont {Stockill}}]{weaver_integrated_2024}%
  \BibitemOpen
  \bibfield  {author} {\bibinfo {author} {\bibfnamefont {M.~J.}\ \bibnamefont {Weaver}}, \bibinfo {author} {\bibfnamefont {P.}~\bibnamefont {Duivestein}}, \bibinfo {author} {\bibfnamefont {A.~C.}\ \bibnamefont {Bernasconi}}, \bibinfo {author} {\bibfnamefont {S.}~\bibnamefont {Scharmer}}, \bibinfo {author} {\bibfnamefont {M.}~\bibnamefont {Lemang}}, \bibinfo {author} {\bibfnamefont {T.~C.}\ \bibnamefont {van Thiel}}, \bibinfo {author} {\bibfnamefont {F.}~\bibnamefont {Hijazi}}, \bibinfo {author} {\bibfnamefont {B.}~\bibnamefont {Hensen}}, \bibinfo {author} {\bibfnamefont {S.}~\bibnamefont {Gr{\"o}blacher}}, \ and\ \bibinfo {author} {\bibfnamefont {R.}~\bibnamefont {Stockill}},\ }\bibfield  {title} {\enquote {\bibinfo {title} {An integrated microwave-to-optics interface for scalable quantum computing},}\ }\href {\doibase 10.1038/s41565-023-01515-y} {\bibfield  {journal} {\bibinfo  {journal} {Nature Nanotechnology}\ }\textbf {\bibinfo {volume} {19}},\ \bibinfo {pages} {166--172} (\bibinfo {year}
  {2024})}\BibitemShut {NoStop}%
\bibitem [{\citenamefont {Zhao}\ \emph {et~al.}(2024)\citenamefont {Zhao}, \citenamefont {Chen}, \citenamefont {Kejriwal},\ and\ \citenamefont {Mirhosseini}}]{zhao_quantum-enabled_2024}%
  \BibitemOpen
  \bibfield  {author} {\bibinfo {author} {\bibfnamefont {H.}~\bibnamefont {Zhao}}, \bibinfo {author} {\bibfnamefont {W.~D.}\ \bibnamefont {Chen}}, \bibinfo {author} {\bibfnamefont {A.}~\bibnamefont {Kejriwal}}, \ and\ \bibinfo {author} {\bibfnamefont {M.}~\bibnamefont {Mirhosseini}},\ }\href@noop {} {\enquote {\bibinfo {title} {Quantum-enabled continuous microwave-to-optics frequency conversion},}\ } (\bibinfo {year} {2024}),\ \Eprint {http://arxiv.org/abs/2406.02704} {arXiv:2406.02704 [physics, physics:quant-ph]} \BibitemShut {NoStop}%
\bibitem [{\citenamefont {Meenehan}\ \emph {et~al.}(2015)\citenamefont {Meenehan}, \citenamefont {Cohen}, \citenamefont {MacCabe}, \citenamefont {Marsili}, \citenamefont {Shaw},\ and\ \citenamefont {Painter}}]{meenehan_pulsed_2015}%
  \BibitemOpen
  \bibfield  {author} {\bibinfo {author} {\bibfnamefont {S.~M.}\ \bibnamefont {Meenehan}}, \bibinfo {author} {\bibfnamefont {J.~D.}\ \bibnamefont {Cohen}}, \bibinfo {author} {\bibfnamefont {G.~S.}\ \bibnamefont {MacCabe}}, \bibinfo {author} {\bibfnamefont {F.}~\bibnamefont {Marsili}}, \bibinfo {author} {\bibfnamefont {M.~D.}\ \bibnamefont {Shaw}}, \ and\ \bibinfo {author} {\bibfnamefont {O.}~\bibnamefont {Painter}},\ }\bibfield  {title} {\enquote {\bibinfo {title} {Pulsed {{Excitation Dynamics}} of an {{Optomechanical Crystal Resonator}} near {{Its Quantum Ground State}} of {{Motion}}},}\ }\href {\doibase 10.1103/PhysRevX.5.041002} {\bibfield  {journal} {\bibinfo  {journal} {Physical Review X}\ }\textbf {\bibinfo {volume} {5}},\ \bibinfo {pages} {041002} (\bibinfo {year} {2015})}\BibitemShut {NoStop}%
\bibitem [{\citenamefont {Ren}\ \emph {et~al.}(2020)\citenamefont {Ren}, \citenamefont {Matheny}, \citenamefont {MacCabe}, \citenamefont {Luo}, \citenamefont {Pfeifer}, \citenamefont {Mirhosseini},\ and\ \citenamefont {Painter}}]{ren_two-dimensional_2020}%
  \BibitemOpen
  \bibfield  {author} {\bibinfo {author} {\bibfnamefont {H.}~\bibnamefont {Ren}}, \bibinfo {author} {\bibfnamefont {M.~H.}\ \bibnamefont {Matheny}}, \bibinfo {author} {\bibfnamefont {G.~S.}\ \bibnamefont {MacCabe}}, \bibinfo {author} {\bibfnamefont {J.}~\bibnamefont {Luo}}, \bibinfo {author} {\bibfnamefont {H.}~\bibnamefont {Pfeifer}}, \bibinfo {author} {\bibfnamefont {M.}~\bibnamefont {Mirhosseini}}, \ and\ \bibinfo {author} {\bibfnamefont {O.}~\bibnamefont {Painter}},\ }\bibfield  {title} {\enquote {\bibinfo {title} {Two-dimensional optomechanical crystal cavity with high quantum cooperativity},}\ }\href {\doibase 10.1038/s41467-020-17182-9} {\bibfield  {journal} {\bibinfo  {journal} {Nature Communications}\ }\textbf {\bibinfo {volume} {11}},\ \bibinfo {pages} {3373} (\bibinfo {year} {2020})}\BibitemShut {NoStop}%
\bibitem [{\citenamefont {Meesala}\ \emph {et~al.}(2024)\citenamefont {Meesala}, \citenamefont {Wood}, \citenamefont {Lake}, \citenamefont {Chiappina}, \citenamefont {Zhong}, \citenamefont {Beyer}, \citenamefont {Shaw}, \citenamefont {Jiang},\ and\ \citenamefont {Painter}}]{meesala_non-classical_2024}%
  \BibitemOpen
  \bibfield  {author} {\bibinfo {author} {\bibfnamefont {S.}~\bibnamefont {Meesala}}, \bibinfo {author} {\bibfnamefont {S.}~\bibnamefont {Wood}}, \bibinfo {author} {\bibfnamefont {D.}~\bibnamefont {Lake}}, \bibinfo {author} {\bibfnamefont {P.}~\bibnamefont {Chiappina}}, \bibinfo {author} {\bibfnamefont {C.}~\bibnamefont {Zhong}}, \bibinfo {author} {\bibfnamefont {A.~D.}\ \bibnamefont {Beyer}}, \bibinfo {author} {\bibfnamefont {M.~D.}\ \bibnamefont {Shaw}}, \bibinfo {author} {\bibfnamefont {L.}~\bibnamefont {Jiang}}, \ and\ \bibinfo {author} {\bibfnamefont {O.}~\bibnamefont {Painter}},\ }\bibfield  {title} {\enquote {\bibinfo {title} {Non-classical microwave--optical photon pair generation with a chip-scale transducer},}\ }\href {\doibase 10.1038/s41567-024-02409-z} {\bibfield  {journal} {\bibinfo  {journal} {Nature Physics}\ }\textbf {\bibinfo {volume} {20}},\ \bibinfo {pages} {871--877} (\bibinfo {year} {2024})}\BibitemShut {NoStop}%
\bibitem [{\citenamefont {{Safavi-Naeini}}\ \emph {et~al.}(2014)\citenamefont {{Safavi-Naeini}}, \citenamefont {Hill}, \citenamefont {Meenehan}, \citenamefont {Chan}, \citenamefont {Gr{\"o}blacher},\ and\ \citenamefont {Painter}}]{safavi-naeini_two-dimensional_2014}%
  \BibitemOpen
  \bibfield  {author} {\bibinfo {author} {\bibfnamefont {A.~H.}\ \bibnamefont {{Safavi-Naeini}}}, \bibinfo {author} {\bibfnamefont {J.~T.}\ \bibnamefont {Hill}}, \bibinfo {author} {\bibfnamefont {S.}~\bibnamefont {Meenehan}}, \bibinfo {author} {\bibfnamefont {J.}~\bibnamefont {Chan}}, \bibinfo {author} {\bibfnamefont {S.}~\bibnamefont {Gr{\"o}blacher}}, \ and\ \bibinfo {author} {\bibfnamefont {O.}~\bibnamefont {Painter}},\ }\bibfield  {title} {\enquote {\bibinfo {title} {Two-{{Dimensional Phononic-Photonic Band Gap Optomechanical Crystal Cavity}}},}\ }\href {\doibase 10.1103/PhysRevLett.112.153603} {\bibfield  {journal} {\bibinfo  {journal} {Physical Review Letters}\ }\textbf {\bibinfo {volume} {112}},\ \bibinfo {pages} {153603} (\bibinfo {year} {2014})}\BibitemShut {NoStop}%
\bibitem [{\citenamefont {Sonar}\ \emph {et~al.}(2024)\citenamefont {Sonar}, \citenamefont {Hatipoglu}, \citenamefont {Meesala}, \citenamefont {Lake}, \citenamefont {Ren},\ and\ \citenamefont {Painter}}]{sonar_high-efficiency_2024}%
  \BibitemOpen
  \bibfield  {author} {\bibinfo {author} {\bibfnamefont {S.}~\bibnamefont {Sonar}}, \bibinfo {author} {\bibfnamefont {U.}~\bibnamefont {Hatipoglu}}, \bibinfo {author} {\bibfnamefont {S.}~\bibnamefont {Meesala}}, \bibinfo {author} {\bibfnamefont {D.}~\bibnamefont {Lake}}, \bibinfo {author} {\bibfnamefont {H.}~\bibnamefont {Ren}}, \ and\ \bibinfo {author} {\bibfnamefont {O.}~\bibnamefont {Painter}},\ }\href@noop {} {\enquote {\bibinfo {title} {High-{{Efficiency Low-Noise Optomechanical Crystal Photon-Phonon Transducers}}},}\ } (\bibinfo {year} {2024}),\ \Eprint {http://arxiv.org/abs/2406.15701} {arXiv:2406.15701 [physics, physics:quant-ph]} \BibitemShut {NoStop}%
\bibitem [{\citenamefont {Mayor}\ \emph {et~al.}(2024)\citenamefont {Mayor}, \citenamefont {Malik}, \citenamefont {Primo}, \citenamefont {Gyger}, \citenamefont {Jiang}, \citenamefont {Alegre},\ and\ \citenamefont {{Safavi-Naeini}}}]{mayor_two-dimensional_2024}%
  \BibitemOpen
  \bibfield  {author} {\bibinfo {author} {\bibfnamefont {F.~M.}\ \bibnamefont {Mayor}}, \bibinfo {author} {\bibfnamefont {S.}~\bibnamefont {Malik}}, \bibinfo {author} {\bibfnamefont {A.~G.}\ \bibnamefont {Primo}}, \bibinfo {author} {\bibfnamefont {S.}~\bibnamefont {Gyger}}, \bibinfo {author} {\bibfnamefont {W.}~\bibnamefont {Jiang}}, \bibinfo {author} {\bibfnamefont {T.~P.~M.}\ \bibnamefont {Alegre}}, \ and\ \bibinfo {author} {\bibfnamefont {A.~H.}\ \bibnamefont {{Safavi-Naeini}}},\ }\href@noop {} {\enquote {\bibinfo {title} {A two-dimensional optomechanical crystal for quantum transduction},}\ } (\bibinfo {year} {2024}),\ \Eprint {http://arxiv.org/abs/2406.14484} {arXiv:2406.14484 [physics, physics:quant-ph]} \BibitemShut {NoStop}%
\bibitem [{\citenamefont {Kolvik}\ \emph {et~al.}(2023)\citenamefont {Kolvik}, \citenamefont {Burger}, \citenamefont {Frey},\ and\ \citenamefont {Van~Laer}}]{kolvik_clamped_2023-1}%
  \BibitemOpen
  \bibfield  {author} {\bibinfo {author} {\bibfnamefont {J.}~\bibnamefont {Kolvik}}, \bibinfo {author} {\bibfnamefont {P.}~\bibnamefont {Burger}}, \bibinfo {author} {\bibfnamefont {J.}~\bibnamefont {Frey}}, \ and\ \bibinfo {author} {\bibfnamefont {R.}~\bibnamefont {Van~Laer}},\ }\bibfield  {title} {\enquote {\bibinfo {title} {Clamped and sideband-resolved silicon optomechanical crystals},}\ }\href {\doibase 10.1364/OPTICA.492143} {\bibfield  {journal} {\bibinfo  {journal} {Optica}\ }\textbf {\bibinfo {volume} {10}},\ \bibinfo {pages} {913--916} (\bibinfo {year} {2023})}\BibitemShut {NoStop}%
\bibitem [{\citenamefont {Van~Laer}, \citenamefont {Baets},\ and\ \citenamefont {Van~Thourhout}(2016)}]{van_laer_unifying_2016}%
  \BibitemOpen
  \bibfield  {author} {\bibinfo {author} {\bibfnamefont {R.}~\bibnamefont {Van~Laer}}, \bibinfo {author} {\bibfnamefont {R.}~\bibnamefont {Baets}}, \ and\ \bibinfo {author} {\bibfnamefont {D.}~\bibnamefont {Van~Thourhout}},\ }\bibfield  {title} {\enquote {\bibinfo {title} {Unifying {{Brillouin}} scattering and cavity optomechanics},}\ }\href {\doibase 10.1103/PhysRevA.93.053828} {\bibfield  {journal} {\bibinfo  {journal} {Physical Review A}\ }\textbf {\bibinfo {volume} {93}},\ \bibinfo {pages} {053828} (\bibinfo {year} {2016})}\BibitemShut {NoStop}%
\bibitem [{\citenamefont {Eggleton}\ \emph {et~al.}(2019)\citenamefont {Eggleton}, \citenamefont {Poulton}, \citenamefont {Rakich}, \citenamefont {Steel},\ and\ \citenamefont {Bahl}}]{eggleton_brillouin_2019}%
  \BibitemOpen
  \bibfield  {author} {\bibinfo {author} {\bibfnamefont {B.~J.}\ \bibnamefont {Eggleton}}, \bibinfo {author} {\bibfnamefont {C.~G.}\ \bibnamefont {Poulton}}, \bibinfo {author} {\bibfnamefont {P.~T.}\ \bibnamefont {Rakich}}, \bibinfo {author} {\bibfnamefont {{\relax Michael}.~J.}\ \bibnamefont {Steel}}, \ and\ \bibinfo {author} {\bibfnamefont {G.}~\bibnamefont {Bahl}},\ }\bibfield  {title} {\enquote {\bibinfo {title} {Brillouin integrated photonics},}\ }\href {\doibase 10.1038/s41566-019-0498-z} {\bibfield  {journal} {\bibinfo  {journal} {Nature Photonics}\ }\textbf {\bibinfo {volume} {13}},\ \bibinfo {pages} {664--677} (\bibinfo {year} {2019})}\BibitemShut {NoStop}%
\bibitem [{\citenamefont {Van~Laer}\ \emph {et~al.}(2015)\citenamefont {Van~Laer}, \citenamefont {Kuyken}, \citenamefont {Van~Thourhout},\ and\ \citenamefont {Baets}}]{van_laer_interaction_2015}%
  \BibitemOpen
  \bibfield  {author} {\bibinfo {author} {\bibfnamefont {R.}~\bibnamefont {Van~Laer}}, \bibinfo {author} {\bibfnamefont {B.}~\bibnamefont {Kuyken}}, \bibinfo {author} {\bibfnamefont {D.}~\bibnamefont {Van~Thourhout}}, \ and\ \bibinfo {author} {\bibfnamefont {R.}~\bibnamefont {Baets}},\ }\bibfield  {title} {\enquote {\bibinfo {title} {Interaction between light and highly confined hypersound in a silicon photonic nanowire},}\ }\href {\doibase 10.1038/nphoton.2015.11} {\bibfield  {journal} {\bibinfo  {journal} {Nature Photonics}\ }\textbf {\bibinfo {volume} {9}},\ \bibinfo {pages} {199--203} (\bibinfo {year} {2015})}\BibitemShut {NoStop}%
\bibitem [{\citenamefont {Qiu}\ \emph {et~al.}(2013)\citenamefont {Qiu}, \citenamefont {Rakich}, \citenamefont {Shin}, \citenamefont {Dong}, \citenamefont {Solja{\v c}i{\'c}},\ and\ \citenamefont {Wang}}]{qiu_stimulated_2013}%
  \BibitemOpen
  \bibfield  {author} {\bibinfo {author} {\bibfnamefont {W.}~\bibnamefont {Qiu}}, \bibinfo {author} {\bibfnamefont {P.~T.}\ \bibnamefont {Rakich}}, \bibinfo {author} {\bibfnamefont {H.}~\bibnamefont {Shin}}, \bibinfo {author} {\bibfnamefont {H.}~\bibnamefont {Dong}}, \bibinfo {author} {\bibfnamefont {M.}~\bibnamefont {Solja{\v c}i{\'c}}}, \ and\ \bibinfo {author} {\bibfnamefont {Z.}~\bibnamefont {Wang}},\ }\bibfield  {title} {\enquote {\bibinfo {title} {Stimulated {{Brillouin}} scattering in nanoscale silicon step-index waveguides: A general framework of selection rules and calculating {{SBS}} gain},}\ }\href {\doibase 10.1364/OE.21.031402} {\bibfield  {journal} {\bibinfo  {journal} {Optics Express}\ }\textbf {\bibinfo {volume} {21}},\ \bibinfo {pages} {31402} (\bibinfo {year} {2013})}\BibitemShut {NoStop}%
\bibitem [{\citenamefont {Place}\ \emph {et~al.}(2021)\citenamefont {Place}, \citenamefont {Rodgers}, \citenamefont {Mundada}, \citenamefont {Smitham}, \citenamefont {Fitzpatrick}, \citenamefont {Leng}, \citenamefont {Premkumar}, \citenamefont {Bryon}, \citenamefont {Vrajitoarea}, \citenamefont {Sussman}, \citenamefont {Cheng}, \citenamefont {Madhavan}, \citenamefont {Babla}, \citenamefont {Le}, \citenamefont {Gang}, \citenamefont {J{\"a}ck}, \citenamefont {Gyenis}, \citenamefont {Yao}, \citenamefont {Cava}, \citenamefont {{de Leon}},\ and\ \citenamefont {Houck}}]{place_new_2021}%
  \BibitemOpen
  \bibfield  {author} {\bibinfo {author} {\bibfnamefont {A.~P.~M.}\ \bibnamefont {Place}}, \bibinfo {author} {\bibfnamefont {L.~V.~H.}\ \bibnamefont {Rodgers}}, \bibinfo {author} {\bibfnamefont {P.}~\bibnamefont {Mundada}}, \bibinfo {author} {\bibfnamefont {B.~M.}\ \bibnamefont {Smitham}}, \bibinfo {author} {\bibfnamefont {M.}~\bibnamefont {Fitzpatrick}}, \bibinfo {author} {\bibfnamefont {Z.}~\bibnamefont {Leng}}, \bibinfo {author} {\bibfnamefont {A.}~\bibnamefont {Premkumar}}, \bibinfo {author} {\bibfnamefont {J.}~\bibnamefont {Bryon}}, \bibinfo {author} {\bibfnamefont {A.}~\bibnamefont {Vrajitoarea}}, \bibinfo {author} {\bibfnamefont {S.}~\bibnamefont {Sussman}}, \bibinfo {author} {\bibfnamefont {G.}~\bibnamefont {Cheng}}, \bibinfo {author} {\bibfnamefont {T.}~\bibnamefont {Madhavan}}, \bibinfo {author} {\bibfnamefont {H.~K.}\ \bibnamefont {Babla}}, \bibinfo {author} {\bibfnamefont {X.~H.}\ \bibnamefont {Le}}, \bibinfo {author} {\bibfnamefont {Y.}~\bibnamefont {Gang}}, \bibinfo {author} {\bibfnamefont
  {B.}~\bibnamefont {J{\"a}ck}}, \bibinfo {author} {\bibfnamefont {A.}~\bibnamefont {Gyenis}}, \bibinfo {author} {\bibfnamefont {N.}~\bibnamefont {Yao}}, \bibinfo {author} {\bibfnamefont {R.~J.}\ \bibnamefont {Cava}}, \bibinfo {author} {\bibfnamefont {N.~P.}\ \bibnamefont {{de Leon}}}, \ and\ \bibinfo {author} {\bibfnamefont {A.~A.}\ \bibnamefont {Houck}},\ }\bibfield  {title} {\enquote {\bibinfo {title} {New material platform for superconducting transmon qubits with coherence times exceeding 0.3 milliseconds},}\ }\href {\doibase 10.1038/s41467-021-22030-5} {\bibfield  {journal} {\bibinfo  {journal} {Nature Communications}\ }\textbf {\bibinfo {volume} {12}},\ \bibinfo {pages} {1779} (\bibinfo {year} {2021})}\BibitemShut {NoStop}%
\bibitem [{\citenamefont {{von L{\"u}pke}}\ \emph {et~al.}(2022)\citenamefont {{von L{\"u}pke}}, \citenamefont {Yang}, \citenamefont {Bild}, \citenamefont {Michaud}, \citenamefont {Fadel},\ and\ \citenamefont {Chu}}]{von_lupke_parity_2022}%
  \BibitemOpen
  \bibfield  {author} {\bibinfo {author} {\bibfnamefont {U.}~\bibnamefont {{von L{\"u}pke}}}, \bibinfo {author} {\bibfnamefont {Y.}~\bibnamefont {Yang}}, \bibinfo {author} {\bibfnamefont {M.}~\bibnamefont {Bild}}, \bibinfo {author} {\bibfnamefont {L.}~\bibnamefont {Michaud}}, \bibinfo {author} {\bibfnamefont {M.}~\bibnamefont {Fadel}}, \ and\ \bibinfo {author} {\bibfnamefont {Y.}~\bibnamefont {Chu}},\ }\bibfield  {title} {\enquote {\bibinfo {title} {Parity measurement in the strong dispersive regime of circuit quantum acoustodynamics},}\ }\href {\doibase 10.1038/s41567-022-01591-2} {\bibfield  {journal} {\bibinfo  {journal} {Nature Physics}\ }\textbf {\bibinfo {volume} {18}},\ \bibinfo {pages} {794--799} (\bibinfo {year} {2022})}\BibitemShut {NoStop}%
\bibitem [{\citenamefont {Martinis}\ and\ \citenamefont {Megrant}(2014)}]{martinis_ucsb_2014}%
  \BibitemOpen
  \bibfield  {author} {\bibinfo {author} {\bibfnamefont {J.~M.}\ \bibnamefont {Martinis}}\ and\ \bibinfo {author} {\bibfnamefont {A.}~\bibnamefont {Megrant}},\ }\href {\doibase 10.48550/arXiv.1410.5793} {\enquote {\bibinfo {title} {{{UCSB}} final report for the {{CSQ}} program: {{Review}} of decoherence and materials physics for superconducting qubits},}\ } (\bibinfo {year} {2014}),\ \Eprint {http://arxiv.org/abs/1410.5793} {arXiv:1410.5793 [cond-mat, physics:quant-ph]} \BibitemShut {NoStop}%
\bibitem [{\citenamefont {{Arrangoiz-Arriola}}\ \emph {et~al.}(2019)\citenamefont {{Arrangoiz-Arriola}}, \citenamefont {Wollack}, \citenamefont {Wang}, \citenamefont {Pechal}, \citenamefont {Jiang}, \citenamefont {McKenna}, \citenamefont {Witmer}, \citenamefont {Van~Laer},\ and\ \citenamefont {{Safavi-Naeini}}}]{arrangoiz-arriola_resolving_2019}%
  \BibitemOpen
  \bibfield  {author} {\bibinfo {author} {\bibfnamefont {P.}~\bibnamefont {{Arrangoiz-Arriola}}}, \bibinfo {author} {\bibfnamefont {E.~A.}\ \bibnamefont {Wollack}}, \bibinfo {author} {\bibfnamefont {Z.}~\bibnamefont {Wang}}, \bibinfo {author} {\bibfnamefont {M.}~\bibnamefont {Pechal}}, \bibinfo {author} {\bibfnamefont {W.}~\bibnamefont {Jiang}}, \bibinfo {author} {\bibfnamefont {T.~P.}\ \bibnamefont {McKenna}}, \bibinfo {author} {\bibfnamefont {J.~D.}\ \bibnamefont {Witmer}}, \bibinfo {author} {\bibfnamefont {R.}~\bibnamefont {Van~Laer}}, \ and\ \bibinfo {author} {\bibfnamefont {A.~H.}\ \bibnamefont {{Safavi-Naeini}}},\ }\bibfield  {title} {\enquote {\bibinfo {title} {Resolving the energy levels of a nanomechanical oscillator},}\ }\href {\doibase 10.1038/s41586-019-1386-x} {\bibfield  {journal} {\bibinfo  {journal} {Nature}\ }\textbf {\bibinfo {volume} {571}},\ \bibinfo {pages} {537--540} (\bibinfo {year} {2019})}\BibitemShut {NoStop}%
\bibitem [{\citenamefont {Wollack}\ \emph {et~al.}(2022)\citenamefont {Wollack}, \citenamefont {Cleland}, \citenamefont {Gruenke}, \citenamefont {Wang}, \citenamefont {{Arrangoiz-Arriola}},\ and\ \citenamefont {{Safavi-Naeini}}}]{wollack_quantum_2022}%
  \BibitemOpen
  \bibfield  {author} {\bibinfo {author} {\bibfnamefont {E.~A.}\ \bibnamefont {Wollack}}, \bibinfo {author} {\bibfnamefont {A.~Y.}\ \bibnamefont {Cleland}}, \bibinfo {author} {\bibfnamefont {R.~G.}\ \bibnamefont {Gruenke}}, \bibinfo {author} {\bibfnamefont {Z.}~\bibnamefont {Wang}}, \bibinfo {author} {\bibfnamefont {P.}~\bibnamefont {{Arrangoiz-Arriola}}}, \ and\ \bibinfo {author} {\bibfnamefont {A.~H.}\ \bibnamefont {{Safavi-Naeini}}},\ }\bibfield  {title} {\enquote {\bibinfo {title} {Quantum state preparation and tomography of entangled mechanical resonators},}\ }\href {\doibase 10.1038/s41586-022-04500-y} {\bibfield  {journal} {\bibinfo  {journal} {Nature}\ }\textbf {\bibinfo {volume} {604}},\ \bibinfo {pages} {463--467} (\bibinfo {year} {2022})}\BibitemShut {NoStop}%
\bibitem [{\citenamefont {MacCabe}\ \emph {et~al.}(2020)\citenamefont {MacCabe}, \citenamefont {Ren}, \citenamefont {Luo}, \citenamefont {Cohen}, \citenamefont {Zhou}, \citenamefont {Sipahigil}, \citenamefont {Mirhosseini},\ and\ \citenamefont {Painter}}]{maccabe_nano-acoustic_2020}%
  \BibitemOpen
  \bibfield  {author} {\bibinfo {author} {\bibfnamefont {G.~S.}\ \bibnamefont {MacCabe}}, \bibinfo {author} {\bibfnamefont {H.}~\bibnamefont {Ren}}, \bibinfo {author} {\bibfnamefont {J.}~\bibnamefont {Luo}}, \bibinfo {author} {\bibfnamefont {J.~D.}\ \bibnamefont {Cohen}}, \bibinfo {author} {\bibfnamefont {H.}~\bibnamefont {Zhou}}, \bibinfo {author} {\bibfnamefont {A.}~\bibnamefont {Sipahigil}}, \bibinfo {author} {\bibfnamefont {M.}~\bibnamefont {Mirhosseini}}, \ and\ \bibinfo {author} {\bibfnamefont {O.}~\bibnamefont {Painter}},\ }\bibfield  {title} {\enquote {\bibinfo {title} {Nano-acoustic resonator with ultralong phonon lifetime},}\ }\href {\doibase 10.1126/science.abc7312} {\bibfield  {journal} {\bibinfo  {journal} {Science}\ }\textbf {\bibinfo {volume} {370}},\ \bibinfo {pages} {840--843} (\bibinfo {year} {2020})}\BibitemShut {NoStop}%
\bibitem [{\citenamefont {Liu}, \citenamefont {Tong},\ and\ \citenamefont {Fang}(2022)}]{liu_optomechanical_2022}%
  \BibitemOpen
  \bibfield  {author} {\bibinfo {author} {\bibfnamefont {S.}~\bibnamefont {Liu}}, \bibinfo {author} {\bibfnamefont {H.}~\bibnamefont {Tong}}, \ and\ \bibinfo {author} {\bibfnamefont {K.}~\bibnamefont {Fang}},\ }\bibfield  {title} {\enquote {\bibinfo {title} {Optomechanical crystal with bound states in the continuum},}\ }\href {\doibase 10.1038/s41467-022-30965-6} {\bibfield  {journal} {\bibinfo  {journal} {Nature Communications}\ }\textbf {\bibinfo {volume} {13}},\ \bibinfo {pages} {3187} (\bibinfo {year} {2022})}\BibitemShut {NoStop}%
\bibitem [{\citenamefont {Sarabalis}\ \emph {et~al.}(2017)\citenamefont {Sarabalis}, \citenamefont {Dahmani}, \citenamefont {Patel}, \citenamefont {Hill},\ and\ \citenamefont {{Safavi-Naeini}}}]{sarabalis_release-free_2017}%
  \BibitemOpen
  \bibfield  {author} {\bibinfo {author} {\bibfnamefont {C.~J.}\ \bibnamefont {Sarabalis}}, \bibinfo {author} {\bibfnamefont {Y.~D.}\ \bibnamefont {Dahmani}}, \bibinfo {author} {\bibfnamefont {R.~N.}\ \bibnamefont {Patel}}, \bibinfo {author} {\bibfnamefont {J.~T.}\ \bibnamefont {Hill}}, \ and\ \bibinfo {author} {\bibfnamefont {A.~H.}\ \bibnamefont {{Safavi-Naeini}}},\ }\bibfield  {title} {\enquote {\bibinfo {title} {Release-free silicon-on-insulator cavity optomechanics},}\ }\href {\doibase 10.1364/OPTICA.4.001147} {\bibfield  {journal} {\bibinfo  {journal} {Optica}\ }\textbf {\bibinfo {volume} {4}},\ \bibinfo {pages} {1147--1150} (\bibinfo {year} {2017})}\BibitemShut {NoStop}%
\bibitem [{\citenamefont {Mason}\ \emph {et~al.}(1971)\citenamefont {Mason}, \citenamefont {de~la Rue}, \citenamefont {Schmidt}, \citenamefont {Ash},\ and\ \citenamefont {Lagasse}}]{mason_ridge_1971}%
  \BibitemOpen
  \bibfield  {author} {\bibinfo {author} {\bibfnamefont {I.~M.}\ \bibnamefont {Mason}}, \bibinfo {author} {\bibfnamefont {R.~M.}\ \bibnamefont {de~la Rue}}, \bibinfo {author} {\bibfnamefont {R.~V.}\ \bibnamefont {Schmidt}}, \bibinfo {author} {\bibfnamefont {E.~A.}\ \bibnamefont {Ash}}, \ and\ \bibinfo {author} {\bibfnamefont {P.~E.}\ \bibnamefont {Lagasse}},\ }\bibfield  {title} {\enquote {\bibinfo {title} {Ridge guides for acoustic surface waves},}\ }\href {\doibase 10.1049/el:19710269} {\bibfield  {journal} {\bibinfo  {journal} {Electronics Letters}\ }\textbf {\bibinfo {volume} {7}},\ \bibinfo {pages} {395--397} (\bibinfo {year} {1971})}\BibitemShut {NoStop}%
\bibitem [{\citenamefont {Lagasse}(1973)}]{lagasse_higherorder_1973}%
  \BibitemOpen
  \bibfield  {author} {\bibinfo {author} {\bibfnamefont {P.~E.}\ \bibnamefont {Lagasse}},\ }\bibfield  {title} {\enquote {\bibinfo {title} {Higher-order finite-element analysis of topographic guides supporting elastic surface waves},}\ }\href {\doibase 10.1121/1.1913432} {\bibfield  {journal} {\bibinfo  {journal} {The Journal of the Acoustical Society of America}\ }\textbf {\bibinfo {volume} {53}},\ \bibinfo {pages} {1116--1122} (\bibinfo {year} {1973})}\BibitemShut {NoStop}%
\bibitem [{\citenamefont {Wollack}\ \emph {et~al.}(2021)\citenamefont {Wollack}, \citenamefont {Cleland}, \citenamefont {{Arrangoiz-Arriola}}, \citenamefont {McKenna}, \citenamefont {Gruenke}, \citenamefont {Patel}, \citenamefont {Jiang}, \citenamefont {Sarabalis},\ and\ \citenamefont {{Safavi-Naeini}}}]{wollack_loss_2021}%
  \BibitemOpen
  \bibfield  {author} {\bibinfo {author} {\bibfnamefont {E.~A.}\ \bibnamefont {Wollack}}, \bibinfo {author} {\bibfnamefont {A.~Y.}\ \bibnamefont {Cleland}}, \bibinfo {author} {\bibfnamefont {P.}~\bibnamefont {{Arrangoiz-Arriola}}}, \bibinfo {author} {\bibfnamefont {T.~P.}\ \bibnamefont {McKenna}}, \bibinfo {author} {\bibfnamefont {R.~G.}\ \bibnamefont {Gruenke}}, \bibinfo {author} {\bibfnamefont {R.~N.}\ \bibnamefont {Patel}}, \bibinfo {author} {\bibfnamefont {W.}~\bibnamefont {Jiang}}, \bibinfo {author} {\bibfnamefont {C.~J.}\ \bibnamefont {Sarabalis}}, \ and\ \bibinfo {author} {\bibfnamefont {A.~H.}\ \bibnamefont {{Safavi-Naeini}}},\ }\bibfield  {title} {\enquote {\bibinfo {title} {Loss channels affecting lithium niobate phononic crystal resonators at cryogenic temperature},}\ }\href {\doibase 10.1063/5.0034909} {\bibfield  {journal} {\bibinfo  {journal} {Applied Physics Letters}\ }\textbf {\bibinfo {volume} {118}},\ \bibinfo {pages} {123501} (\bibinfo {year} {2021})}\BibitemShut {NoStop}%
\bibitem [{\citenamefont {McKenna}\ \emph {et~al.}(2020)\citenamefont {McKenna}, \citenamefont {Witmer}, \citenamefont {Patel}, \citenamefont {Jiang}, \citenamefont {Van~Laer}, \citenamefont {{Arrangoiz-Arriola}}, \citenamefont {Wollack}, \citenamefont {Herrmann},\ and\ \citenamefont {{Safavi-Naeini}}}]{mckenna_cryogenic_2020}%
  \BibitemOpen
  \bibfield  {author} {\bibinfo {author} {\bibfnamefont {T.~P.}\ \bibnamefont {McKenna}}, \bibinfo {author} {\bibfnamefont {J.~D.}\ \bibnamefont {Witmer}}, \bibinfo {author} {\bibfnamefont {R.~N.}\ \bibnamefont {Patel}}, \bibinfo {author} {\bibfnamefont {W.}~\bibnamefont {Jiang}}, \bibinfo {author} {\bibfnamefont {R.}~\bibnamefont {Van~Laer}}, \bibinfo {author} {\bibfnamefont {P.}~\bibnamefont {{Arrangoiz-Arriola}}}, \bibinfo {author} {\bibfnamefont {E.~A.}\ \bibnamefont {Wollack}}, \bibinfo {author} {\bibfnamefont {J.~F.}\ \bibnamefont {Herrmann}}, \ and\ \bibinfo {author} {\bibfnamefont {A.~H.}\ \bibnamefont {{Safavi-Naeini}}},\ }\bibfield  {title} {\enquote {\bibinfo {title} {Cryogenic microwave-to-optical conversion using a triply resonant lithium-niobate-on-sapphire transducer},}\ }\href {\doibase 10.1364/OPTICA.397235} {\bibfield  {journal} {\bibinfo  {journal} {Optica}\ }\textbf {\bibinfo {volume} {7}},\ \bibinfo {pages} {1737--1745} (\bibinfo {year} {2020})}\BibitemShut {NoStop}%
\bibitem [{\citenamefont {Zorzetti}\ \emph {et~al.}(2023)\citenamefont {Zorzetti}, \citenamefont {Wang}, \citenamefont {Gonin}, \citenamefont {Kazakov}, \citenamefont {Khabiboulline}, \citenamefont {Romanenko}, \citenamefont {Yakovlev},\ and\ \citenamefont {Grassellino}}]{zorzetti_millikelvin_2023}%
  \BibitemOpen
  \bibfield  {author} {\bibinfo {author} {\bibfnamefont {S.}~\bibnamefont {Zorzetti}}, \bibinfo {author} {\bibfnamefont {C.}~\bibnamefont {Wang}}, \bibinfo {author} {\bibfnamefont {I.}~\bibnamefont {Gonin}}, \bibinfo {author} {\bibfnamefont {S.}~\bibnamefont {Kazakov}}, \bibinfo {author} {\bibfnamefont {T.}~\bibnamefont {Khabiboulline}}, \bibinfo {author} {\bibfnamefont {A.}~\bibnamefont {Romanenko}}, \bibinfo {author} {\bibfnamefont {V.~P.}\ \bibnamefont {Yakovlev}}, \ and\ \bibinfo {author} {\bibfnamefont {A.}~\bibnamefont {Grassellino}},\ }\bibfield  {title} {\enquote {\bibinfo {title} {Millikelvin measurements of permittivity and loss tangent of lithium niobate},}\ }\href {\doibase 10.1103/PhysRevB.107.L220302} {\bibfield  {journal} {\bibinfo  {journal} {Physical Review B}\ }\textbf {\bibinfo {volume} {107}},\ \bibinfo {pages} {L220302} (\bibinfo {year} {2023})}\BibitemShut {NoStop}%
\bibitem [{\citenamefont {Chiappina}\ \emph {et~al.}(2023)\citenamefont {Chiappina}, \citenamefont {Banker}, \citenamefont {Meesala}, \citenamefont {Lake}, \citenamefont {Wood},\ and\ \citenamefont {Painter}}]{chiappina_design_2023-1}%
  \BibitemOpen
  \bibfield  {author} {\bibinfo {author} {\bibfnamefont {P.}~\bibnamefont {Chiappina}}, \bibinfo {author} {\bibfnamefont {J.}~\bibnamefont {Banker}}, \bibinfo {author} {\bibfnamefont {S.}~\bibnamefont {Meesala}}, \bibinfo {author} {\bibfnamefont {D.}~\bibnamefont {Lake}}, \bibinfo {author} {\bibfnamefont {S.}~\bibnamefont {Wood}}, \ and\ \bibinfo {author} {\bibfnamefont {O.}~\bibnamefont {Painter}},\ }\bibfield  {title} {\enquote {\bibinfo {title} {Design of an ultra-low mode volume piezo-optomechanical quantum transducer},}\ }\href {\doibase 10.1364/OE.493532} {\bibfield  {journal} {\bibinfo  {journal} {Optics Express}\ }\textbf {\bibinfo {volume} {31}},\ \bibinfo {pages} {22914} (\bibinfo {year} {2023})}\BibitemShut {NoStop}%
\bibitem [{\citenamefont {Chan}(2012)}]{chan_laser_nodate}%
  \BibitemOpen
  \bibfield  {author} {\bibinfo {author} {\bibfnamefont {J.}~\bibnamefont {Chan}},\ }\emph {\bibinfo {title} {Laser Cooling of an Optomechanical Crystal Resonator to Its Quantum Ground State of Motion}},\ \href@noop {} {Ph.D. thesis},\ \bibinfo  {school} {California Institute of Technology} (\bibinfo {year} {2012})\BibitemShut {NoStop}%
\bibitem [{\citenamefont {Wang}\ and\ \citenamefont {Clerk}(2012)}]{wang_using_2012}%
  \BibitemOpen
  \bibfield  {author} {\bibinfo {author} {\bibfnamefont {Y.-D.}\ \bibnamefont {Wang}}\ and\ \bibinfo {author} {\bibfnamefont {A.~A.}\ \bibnamefont {Clerk}},\ }\bibfield  {title} {\enquote {\bibinfo {title} {Using dark modes for high-fidelity optomechanical quantum state transfer},}\ }\href {\doibase 10.1088/1367-2630/14/10/105010} {\bibfield  {journal} {\bibinfo  {journal} {New Journal of Physics}\ }\textbf {\bibinfo {volume} {14}},\ \bibinfo {pages} {105010} (\bibinfo {year} {2012})}\BibitemShut {NoStop}%
\bibitem [{\citenamefont {Jiang}\ \emph {et~al.}(2020)\citenamefont {Jiang}, \citenamefont {Sarabalis}, \citenamefont {Dahmani}, \citenamefont {Patel}, \citenamefont {Mayor}, \citenamefont {McKenna}, \citenamefont {Van~Laer},\ and\ \citenamefont {{Safavi-Naeini}}}]{jiang_efficient_2020}%
  \BibitemOpen
  \bibfield  {author} {\bibinfo {author} {\bibfnamefont {W.}~\bibnamefont {Jiang}}, \bibinfo {author} {\bibfnamefont {C.~J.}\ \bibnamefont {Sarabalis}}, \bibinfo {author} {\bibfnamefont {Y.~D.}\ \bibnamefont {Dahmani}}, \bibinfo {author} {\bibfnamefont {R.~N.}\ \bibnamefont {Patel}}, \bibinfo {author} {\bibfnamefont {F.~M.}\ \bibnamefont {Mayor}}, \bibinfo {author} {\bibfnamefont {T.~P.}\ \bibnamefont {McKenna}}, \bibinfo {author} {\bibfnamefont {R.}~\bibnamefont {Van~Laer}}, \ and\ \bibinfo {author} {\bibfnamefont {A.~H.}\ \bibnamefont {{Safavi-Naeini}}},\ }\bibfield  {title} {\enquote {\bibinfo {title} {Efficient bidirectional piezo-optomechanical transduction between microwave and optical frequency},}\ }\href {\doibase 10.1038/s41467-020-14863-3} {\bibfield  {journal} {\bibinfo  {journal} {Nature Communications}\ }\textbf {\bibinfo {volume} {11}},\ \bibinfo {pages} {1166} (\bibinfo {year} {2020})}\BibitemShut {NoStop}%
\bibitem [{\citenamefont {Marpaung}, \citenamefont {Yao},\ and\ \citenamefont {Capmany}(2019)}]{marpaung_integrated_2019}%
  \BibitemOpen
  \bibfield  {author} {\bibinfo {author} {\bibfnamefont {D.}~\bibnamefont {Marpaung}}, \bibinfo {author} {\bibfnamefont {J.}~\bibnamefont {Yao}}, \ and\ \bibinfo {author} {\bibfnamefont {J.}~\bibnamefont {Capmany}},\ }\bibfield  {title} {\enquote {\bibinfo {title} {Integrated microwave photonics},}\ }\href {\doibase 10.1038/s41566-018-0310-5} {\bibfield  {journal} {\bibinfo  {journal} {Nature Photonics}\ }\textbf {\bibinfo {volume} {13}},\ \bibinfo {pages} {80--90} (\bibinfo {year} {2019})}\BibitemShut {NoStop}%
\bibitem [{\citenamefont {Zhou}\ \emph {et~al.}(2024)\citenamefont {Zhou}, \citenamefont {Ruesink}, \citenamefont {Pavlovich}, \citenamefont {Behunin}, \citenamefont {Cheng}, \citenamefont {Gertler}, \citenamefont {Starbuck}, \citenamefont {Leenheer}, \citenamefont {Pomerene}, \citenamefont {Trotter}, \citenamefont {Musick}, \citenamefont {Gehl}, \citenamefont {Kodigala}, \citenamefont {Eichenfield}, \citenamefont {Lentine}, \citenamefont {Otterstrom},\ and\ \citenamefont {Rakich}}]{zhou_electrically_2024}%
  \BibitemOpen
  \bibfield  {author} {\bibinfo {author} {\bibfnamefont {Y.}~\bibnamefont {Zhou}}, \bibinfo {author} {\bibfnamefont {F.}~\bibnamefont {Ruesink}}, \bibinfo {author} {\bibfnamefont {M.}~\bibnamefont {Pavlovich}}, \bibinfo {author} {\bibfnamefont {R.}~\bibnamefont {Behunin}}, \bibinfo {author} {\bibfnamefont {H.}~\bibnamefont {Cheng}}, \bibinfo {author} {\bibfnamefont {S.}~\bibnamefont {Gertler}}, \bibinfo {author} {\bibfnamefont {A.~L.}\ \bibnamefont {Starbuck}}, \bibinfo {author} {\bibfnamefont {A.~J.}\ \bibnamefont {Leenheer}}, \bibinfo {author} {\bibfnamefont {A.~T.}\ \bibnamefont {Pomerene}}, \bibinfo {author} {\bibfnamefont {D.~C.}\ \bibnamefont {Trotter}}, \bibinfo {author} {\bibfnamefont {K.~M.}\ \bibnamefont {Musick}}, \bibinfo {author} {\bibfnamefont {M.}~\bibnamefont {Gehl}}, \bibinfo {author} {\bibfnamefont {A.}~\bibnamefont {Kodigala}}, \bibinfo {author} {\bibfnamefont {M.}~\bibnamefont {Eichenfield}}, \bibinfo {author} {\bibfnamefont {A.~L.}\ \bibnamefont {Lentine}}, \bibinfo {author}
  {\bibfnamefont {N.}~\bibnamefont {Otterstrom}}, \ and\ \bibinfo {author} {\bibfnamefont {P.}~\bibnamefont {Rakich}},\ }\bibfield  {title} {\enquote {\bibinfo {title} {Electrically interfaced {{Brillouin-active}} waveguide for microwave photonic measurements},}\ }\href {\doibase 10.1038/s41467-024-51010-8} {\bibfield  {journal} {\bibinfo  {journal} {Nature Communications}\ }\textbf {\bibinfo {volume} {15}},\ \bibinfo {pages} {6796} (\bibinfo {year} {2024})}\BibitemShut {NoStop}%
\bibitem [{\citenamefont {Sarabalis}, \citenamefont {Van~Laer},\ and\ \citenamefont {{Safavi-Naeini}}(2018)}]{sarabalis_optomechanical_2018}%
  \BibitemOpen
  \bibfield  {author} {\bibinfo {author} {\bibfnamefont {C.~J.}\ \bibnamefont {Sarabalis}}, \bibinfo {author} {\bibfnamefont {R.}~\bibnamefont {Van~Laer}}, \ and\ \bibinfo {author} {\bibfnamefont {A.~H.}\ \bibnamefont {{Safavi-Naeini}}},\ }\bibfield  {title} {\enquote {\bibinfo {title} {Optomechanical antennas for on-chip beam-steering},}\ }\href {\doibase 10.1364/OE.26.022075} {\bibfield  {journal} {\bibinfo  {journal} {Optics Express}\ }\textbf {\bibinfo {volume} {26}},\ \bibinfo {pages} {22075--22099} (\bibinfo {year} {2018})}\BibitemShut {NoStop}%
\bibitem [{\citenamefont {Zhang}\ \emph {et~al.}(2024)\citenamefont {Zhang}, \citenamefont {Cui}, \citenamefont {Chen},\ and\ \citenamefont {Fan}}]{zhang_integrated-waveguide-based_2024}%
  \BibitemOpen
  \bibfield  {author} {\bibinfo {author} {\bibfnamefont {L.}~\bibnamefont {Zhang}}, \bibinfo {author} {\bibfnamefont {C.}~\bibnamefont {Cui}}, \bibinfo {author} {\bibfnamefont {P.-K.}\ \bibnamefont {Chen}}, \ and\ \bibinfo {author} {\bibfnamefont {L.}~\bibnamefont {Fan}},\ }\bibfield  {title} {\enquote {\bibinfo {title} {Integrated-waveguide-based acousto-optic modulation with complete optical conversion},}\ }\href {\doibase 10.1364/OPTICA.488271} {\bibfield  {journal} {\bibinfo  {journal} {Optica}\ }\textbf {\bibinfo {volume} {11}},\ \bibinfo {pages} {184--189} (\bibinfo {year} {2024})}\BibitemShut {NoStop}%
\bibitem [{\citenamefont {Weis}\ and\ \citenamefont {Gaylord}(1985)}]{weis_lithium_1985}%
  \BibitemOpen
  \bibfield  {author} {\bibinfo {author} {\bibfnamefont {R.~S.}\ \bibnamefont {Weis}}\ and\ \bibinfo {author} {\bibfnamefont {T.~K.}\ \bibnamefont {Gaylord}},\ }\bibfield  {title} {\enquote {\bibinfo {title} {Lithium niobate: {{Summary}} of physical properties and crystal structure},}\ }\href {\doibase 10.1007/BF00614817} {\bibfield  {journal} {\bibinfo  {journal} {Applied Physics A Solids and Surfaces}\ }\textbf {\bibinfo {volume} {37}},\ \bibinfo {pages} {191--203} (\bibinfo {year} {1985})}\BibitemShut {NoStop}%
\bibitem [{\citenamefont {Vodenitcharova}\ \emph {et~al.}(2007)\citenamefont {Vodenitcharova}, \citenamefont {Zhang}, \citenamefont {Zarudi}, \citenamefont {Yin}, \citenamefont {Domyo}, \citenamefont {Ho},\ and\ \citenamefont {Sato}}]{vodenitcharova_effect_2007}%
  \BibitemOpen
  \bibfield  {author} {\bibinfo {author} {\bibfnamefont {T.}~\bibnamefont {Vodenitcharova}}, \bibinfo {author} {\bibfnamefont {L.}~\bibnamefont {Zhang}}, \bibinfo {author} {\bibfnamefont {I.}~\bibnamefont {Zarudi}}, \bibinfo {author} {\bibfnamefont {Y.}~\bibnamefont {Yin}}, \bibinfo {author} {\bibfnamefont {H.}~\bibnamefont {Domyo}}, \bibinfo {author} {\bibfnamefont {T.}~\bibnamefont {Ho}}, \ and\ \bibinfo {author} {\bibfnamefont {M.}~\bibnamefont {Sato}},\ }\bibfield  {title} {\enquote {\bibinfo {title} {The effect of anisotropy on the deformation and fracture of sapphire wafers subjected to thermal shocks},}\ }\href {\doibase 10.1016/j.jmatprotec.2007.03.125} {\bibfield  {journal} {\bibinfo  {journal} {Journal of Materials Processing Technology}\ }\textbf {\bibinfo {volume} {194}},\ \bibinfo {pages} {52--62} (\bibinfo {year} {2007})}\BibitemShut {NoStop}%
\bibitem [{\citenamefont {Auld}(1990)}]{auld_acoustic_1990}%
  \BibitemOpen
  \bibfield  {author} {\bibinfo {author} {\bibfnamefont {B.~A.}\ \bibnamefont {Auld}},\ }\href@noop {} {\emph {\bibinfo {title} {Acoustic Fields and Waves in Solids. 1}}},\ \bibinfo {edition} {2nd}\ ed.\ (\bibinfo {address} {Malabar, Fla},\ \bibinfo {year} {1990})\BibitemShut {NoStop}%
\bibitem [{\citenamefont {Coufal}\ \emph {et~al.}(1994)\citenamefont {Coufal}, \citenamefont {Meyer}, \citenamefont {Grygier}, \citenamefont {Hess},\ and\ \citenamefont {Neubrand}}]{coufal_precision_1994}%
  \BibitemOpen
  \bibfield  {author} {\bibinfo {author} {\bibfnamefont {H.}~\bibnamefont {Coufal}}, \bibinfo {author} {\bibfnamefont {K.}~\bibnamefont {Meyer}}, \bibinfo {author} {\bibfnamefont {R.~K.}\ \bibnamefont {Grygier}}, \bibinfo {author} {\bibfnamefont {P.}~\bibnamefont {Hess}}, \ and\ \bibinfo {author} {\bibfnamefont {A.}~\bibnamefont {Neubrand}},\ }\bibfield  {title} {\enquote {\bibinfo {title} {Precision measurement of the surface acoustic wave velocity on silicon single crystals using optical excitation and detection},}\ }\href {\doibase 10.1121/1.408473} {\bibfield  {journal} {\bibinfo  {journal} {The Journal of the Acoustical Society of America}\ }\textbf {\bibinfo {volume} {95}},\ \bibinfo {pages} {1158--1160} (\bibinfo {year} {1994})}\BibitemShut {NoStop}%
\bibitem [{\citenamefont {Tarasenko}, \citenamefont {{\v C}tvrtl{\'i}k},\ and\ \citenamefont {Kud{\v e}lka}(2021)}]{tarasenko_theoretical_2021}%
  \BibitemOpen
  \bibfield  {author} {\bibinfo {author} {\bibfnamefont {A.}~\bibnamefont {Tarasenko}}, \bibinfo {author} {\bibfnamefont {R.}~\bibnamefont {{\v C}tvrtl{\'i}k}}, \ and\ \bibinfo {author} {\bibfnamefont {R.}~\bibnamefont {Kud{\v e}lka}},\ }\bibfield  {title} {\enquote {\bibinfo {title} {Theoretical and experimental revision of surface acoustic waves on the (100) plane of silicon},}\ }\href {\doibase 10.1038/s41598-021-82211-6} {\bibfield  {journal} {\bibinfo  {journal} {Scientific Reports}\ }\textbf {\bibinfo {volume} {11}},\ \bibinfo {pages} {2845} (\bibinfo {year} {2021})}\BibitemShut {NoStop}%
\bibitem [{\citenamefont {Wolff}\ \emph {et~al.}(2016)\citenamefont {Wolff}, \citenamefont {Van~Laer}, \citenamefont {Steel}, \citenamefont {Eggleton},\ and\ \citenamefont {Poulton}}]{wolff_brillouin_2016}%
  \BibitemOpen
  \bibfield  {author} {\bibinfo {author} {\bibfnamefont {C.}~\bibnamefont {Wolff}}, \bibinfo {author} {\bibfnamefont {R.}~\bibnamefont {Van~Laer}}, \bibinfo {author} {\bibfnamefont {M.~J.}\ \bibnamefont {Steel}}, \bibinfo {author} {\bibfnamefont {B.~J.}\ \bibnamefont {Eggleton}}, \ and\ \bibinfo {author} {\bibfnamefont {C.~G.}\ \bibnamefont {Poulton}},\ }\bibfield  {title} {\enquote {\bibinfo {title} {Brillouin resonance broadening due to structural variations in nanoscale waveguides},}\ }\href {\doibase 10.1088/1367-2630/18/2/025006} {\bibfield  {journal} {\bibinfo  {journal} {New Journal of Physics}\ }\textbf {\bibinfo {volume} {18}},\ \bibinfo {pages} {025006} (\bibinfo {year} {2016})}\BibitemShut {NoStop}%
\bibitem [{\citenamefont {Chan}\ \emph {et~al.}(2009)\citenamefont {Chan}, \citenamefont {Eichenfield}, \citenamefont {Camacho},\ and\ \citenamefont {Painter}}]{chan_optical_2009-1}%
  \BibitemOpen
  \bibfield  {author} {\bibinfo {author} {\bibfnamefont {J.}~\bibnamefont {Chan}}, \bibinfo {author} {\bibfnamefont {M.}~\bibnamefont {Eichenfield}}, \bibinfo {author} {\bibfnamefont {R.}~\bibnamefont {Camacho}}, \ and\ \bibinfo {author} {\bibfnamefont {O.}~\bibnamefont {Painter}},\ }\bibfield  {title} {\enquote {\bibinfo {title} {Optical and mechanical design of a ``zipper'' photonic crystal optomechanical cavity},}\ }\href {\doibase 10.1364/OE.17.003802} {\bibfield  {journal} {\bibinfo  {journal} {Optics Express}\ }\textbf {\bibinfo {volume} {17}},\ \bibinfo {pages} {3802--3817} (\bibinfo {year} {2009})}\BibitemShut {NoStop}%
\bibitem [{\citenamefont {Bl{\'e}sin}\ \emph {et~al.}(2021)\citenamefont {Bl{\'e}sin}, \citenamefont {Tian}, \citenamefont {Bhave},\ and\ \citenamefont {Kippenberg}}]{blesin_quantum_2021}%
  \BibitemOpen
  \bibfield  {author} {\bibinfo {author} {\bibfnamefont {T.}~\bibnamefont {Bl{\'e}sin}}, \bibinfo {author} {\bibfnamefont {H.}~\bibnamefont {Tian}}, \bibinfo {author} {\bibfnamefont {S.~A.}\ \bibnamefont {Bhave}}, \ and\ \bibinfo {author} {\bibfnamefont {T.~J.}\ \bibnamefont {Kippenberg}},\ }\bibfield  {title} {\enquote {\bibinfo {title} {Quantum coherent microwave-optical transduction using high-overtone bulk acoustic resonances},}\ }\href {\doibase 10.1103/PhysRevA.104.052601} {\bibfield  {journal} {\bibinfo  {journal} {Physical Review A}\ }\textbf {\bibinfo {volume} {104}},\ \bibinfo {pages} {052601} (\bibinfo {year} {2021})}\BibitemShut {NoStop}%
\bibitem [{\citenamefont {{Arrangoiz-Arriola}}\ and\ \citenamefont {{Safavi-Naeini}}(2016)}]{arrangoiz-arriola_engineering_2016}%
  \BibitemOpen
  \bibfield  {author} {\bibinfo {author} {\bibfnamefont {P.}~\bibnamefont {{Arrangoiz-Arriola}}}\ and\ \bibinfo {author} {\bibfnamefont {A.~H.}\ \bibnamefont {{Safavi-Naeini}}},\ }\bibfield  {title} {\enquote {\bibinfo {title} {Engineering interactions between superconducting qubits and phononic nanostructures},}\ }\href {\doibase 10.1103/PhysRevA.94.063864} {\bibfield  {journal} {\bibinfo  {journal} {Physical Review A}\ }\textbf {\bibinfo {volume} {94}},\ \bibinfo {pages} {063864} (\bibinfo {year} {2016})}\BibitemShut {NoStop}%
\bibitem [{\citenamefont {Banker}(2022)}]{banker_photonic_nodate}%
  \BibitemOpen
  \bibfield  {author} {\bibinfo {author} {\bibfnamefont {J.~H.}\ \bibnamefont {Banker}},\ }\emph {\bibinfo {title} {Photonic and {{Phononic Band Gap Engineering}} for {{Circuit Quantum Electrodynamics}} and {{Quantum Transduction}}}},\ \href@noop {} {Ph.D. thesis},\ \bibinfo  {school} {California Institute of Technology} (\bibinfo {year} {2022})\BibitemShut {NoStop}%
\bibitem [{\citenamefont {Niepce}, \citenamefont {Burnett},\ and\ \citenamefont {Bylander}(2019)}]{niepce_high_2019}%
  \BibitemOpen
  \bibfield  {author} {\bibinfo {author} {\bibfnamefont {D.}~\bibnamefont {Niepce}}, \bibinfo {author} {\bibfnamefont {J.}~\bibnamefont {Burnett}}, \ and\ \bibinfo {author} {\bibfnamefont {J.}~\bibnamefont {Bylander}},\ }\bibfield  {title} {\enquote {\bibinfo {title} {High {{Kinetic Inductance NbN Nanowire Superinductors}}},}\ }\href {\doibase 10.1103/PhysRevApplied.11.044014} {\bibfield  {journal} {\bibinfo  {journal} {Physical Review Applied}\ }\textbf {\bibinfo {volume} {11}},\ \bibinfo {pages} {044014} (\bibinfo {year} {2019})}\BibitemShut {NoStop}%
\bibitem [{\citenamefont {Luiten}, \citenamefont {Mann},\ and\ \citenamefont {Blair}(1993)}]{luiten_ultrahigh_1993}%
  \BibitemOpen
  \bibfield  {author} {\bibinfo {author} {\bibfnamefont {A.}~\bibnamefont {Luiten}}, \bibinfo {author} {\bibfnamefont {A.}~\bibnamefont {Mann}}, \ and\ \bibinfo {author} {\bibfnamefont {D.}~\bibnamefont {Blair}},\ }\bibfield  {title} {\enquote {\bibinfo {title} {Ultrahigh {{Q-factor}} cryogenic sapphire resonator},}\ }\href {\doibase 10.1049/el:19930587} {\bibfield  {journal} {\bibinfo  {journal} {Electronics Letters}\ }\textbf {\bibinfo {volume} {29}},\ \bibinfo {pages} {879--881} (\bibinfo {year} {1993})}\BibitemShut {NoStop}%
\bibitem [{\citenamefont {Yariv}\ and\ \citenamefont {Yeh}(2003)}]{yariv_optical_2003}%
  \BibitemOpen
  \bibfield  {author} {\bibinfo {author} {\bibfnamefont {A.}~\bibnamefont {Yariv}}\ and\ \bibinfo {author} {\bibfnamefont {P.}~\bibnamefont {Yeh}},\ }\href@noop {} {\emph {\bibinfo {title} {Optical Waves in Crystals: Propagation and Control of Laser Radiation}}},\ \bibinfo {edition} {wiley classics library ed}\ ed.,\ Wiley Classics Library\ (\bibinfo {address} {Hoboken, N.J},\ \bibinfo {year} {2003})\BibitemShut {NoStop}%
\bibitem [{\citenamefont {Schmidt}, \citenamefont {Baker},\ and\ \citenamefont {Van~Laer}(2022)}]{schmidt_convergence_2022}%
  \BibitemOpen
  \bibfield  {author} {\bibinfo {author} {\bibfnamefont {M.~K.}\ \bibnamefont {Schmidt}}, \bibinfo {author} {\bibfnamefont {C.~G.}\ \bibnamefont {Baker}}, \ and\ \bibinfo {author} {\bibfnamefont {R.}~\bibnamefont {Van~Laer}},\ }\bibfield  {title} {\enquote {\bibinfo {title} {The convergence of cavity optomechanics and {{Brillouin}} scattering},}\ }in\ \href {\doibase 10.1016/bs.semsem.2022.04.005} {\emph {\bibinfo {booktitle} {Semiconductors and {{Semimetals}}}}},\ Vol.\ \bibinfo {volume} {109}\ (\bibinfo {year} {2022})\ pp.\ \bibinfo {pages} {93--131}\BibitemShut {NoStop}%
\bibitem [{\citenamefont {Wolff}\ \emph {et~al.}(2022)\citenamefont {Wolff}, \citenamefont {Poulton}, \citenamefont {Steel},\ and\ \citenamefont {Wiederhecker}}]{wolff_chapter_2022}%
  \BibitemOpen
  \bibfield  {author} {\bibinfo {author} {\bibfnamefont {C.}~\bibnamefont {Wolff}}, \bibinfo {author} {\bibfnamefont {C.~G.}\ \bibnamefont {Poulton}}, \bibinfo {author} {\bibfnamefont {M.~J.}\ \bibnamefont {Steel}}, \ and\ \bibinfo {author} {\bibfnamefont {G.}~\bibnamefont {Wiederhecker}},\ }\bibfield  {title} {\enquote {\bibinfo {title} {Chapter {{Two}} - {{Theoretical}} formalisms for stimulated {{Brillouin}} scattering},}\ }in\ \href {\doibase 10.1016/bs.semsem.2022.04.002} {\emph {\bibinfo {booktitle} {Semiconductors and {{Semimetals}}}}},\ \bibinfo {series} {Brillouin {{Scattering Part}} 1}, Vol.\ \bibinfo {volume} {109},\ \bibinfo {editor} {edited by\ \bibinfo {editor} {\bibfnamefont {B.~J.}\ \bibnamefont {Eggleton}}, \bibinfo {editor} {\bibfnamefont {M.~J.}\ \bibnamefont {Steel}}, \ and\ \bibinfo {editor} {\bibfnamefont {C.~G.}\ \bibnamefont {Poulton}}}\ (\bibinfo {year} {2022})\ pp.\ \bibinfo {pages} {27--91}\BibitemShut {NoStop}%
\bibitem [{\citenamefont {Johnson}\ \emph {et~al.}(2002)\citenamefont {Johnson}, \citenamefont {Ibanescu}, \citenamefont {Skorobogatiy}, \citenamefont {Weisberg}, \citenamefont {Joannopoulos},\ and\ \citenamefont {Fink}}]{johnson_perturbation_2002}%
  \BibitemOpen
  \bibfield  {author} {\bibinfo {author} {\bibfnamefont {S.~G.}\ \bibnamefont {Johnson}}, \bibinfo {author} {\bibfnamefont {M.}~\bibnamefont {Ibanescu}}, \bibinfo {author} {\bibfnamefont {M.~A.}\ \bibnamefont {Skorobogatiy}}, \bibinfo {author} {\bibfnamefont {O.}~\bibnamefont {Weisberg}}, \bibinfo {author} {\bibfnamefont {J.~D.}\ \bibnamefont {Joannopoulos}}, \ and\ \bibinfo {author} {\bibfnamefont {Y.}~\bibnamefont {Fink}},\ }\bibfield  {title} {\enquote {\bibinfo {title} {Perturbation theory for {{Maxwell}}'s equations with shifting material boundaries},}\ }\href {\doibase 10.1103/PhysRevE.65.066611} {\bibfield  {journal} {\bibinfo  {journal} {Physical Review E}\ }\textbf {\bibinfo {volume} {65}},\ \bibinfo {pages} {066611} (\bibinfo {year} {2002})}\BibitemShut {NoStop}%
\bibitem [{\citenamefont {Qiu}\ \emph {et~al.}(2012)\citenamefont {Qiu}, \citenamefont {Rakich}, \citenamefont {Soljacic},\ and\ \citenamefont {Wang}}]{qiu_stimulated_2012}%
  \BibitemOpen
  \bibfield  {author} {\bibinfo {author} {\bibfnamefont {W.}~\bibnamefont {Qiu}}, \bibinfo {author} {\bibfnamefont {P.~T.}\ \bibnamefont {Rakich}}, \bibinfo {author} {\bibfnamefont {M.}~\bibnamefont {Soljacic}}, \ and\ \bibinfo {author} {\bibfnamefont {Z.}~\bibnamefont {Wang}},\ }\href@noop {} {\enquote {\bibinfo {title} {Stimulated brillouin scattering in slow light waveguides},}\ } (\bibinfo {year} {2012}),\ \Eprint {http://arxiv.org/abs/1210.0738} {arXiv:1210.0738 [cond-mat, physics:physics]} \BibitemShut {NoStop}%
\bibitem [{\citenamefont {Chan}\ \emph {et~al.}(2011)\citenamefont {Chan}, \citenamefont {Alegre}, \citenamefont {{Safavi-Naeini}}, \citenamefont {Hill}, \citenamefont {Krause}, \citenamefont {Gr{\"o}blacher}, \citenamefont {Aspelmeyer},\ and\ \citenamefont {Painter}}]{chan_laser_2011}%
  \BibitemOpen
  \bibfield  {author} {\bibinfo {author} {\bibfnamefont {J.}~\bibnamefont {Chan}}, \bibinfo {author} {\bibfnamefont {T.~P.~M.}\ \bibnamefont {Alegre}}, \bibinfo {author} {\bibfnamefont {A.~H.}\ \bibnamefont {{Safavi-Naeini}}}, \bibinfo {author} {\bibfnamefont {J.~T.}\ \bibnamefont {Hill}}, \bibinfo {author} {\bibfnamefont {A.}~\bibnamefont {Krause}}, \bibinfo {author} {\bibfnamefont {S.}~\bibnamefont {Gr{\"o}blacher}}, \bibinfo {author} {\bibfnamefont {M.}~\bibnamefont {Aspelmeyer}}, \ and\ \bibinfo {author} {\bibfnamefont {O.}~\bibnamefont {Painter}},\ }\bibfield  {title} {\enquote {\bibinfo {title} {Laser cooling of a nanomechanical oscillator into its quantum ground state},}\ }\href {\doibase 10.1038/nature10461} {\bibfield  {journal} {\bibinfo  {journal} {Nature}\ }\textbf {\bibinfo {volume} {478}},\ \bibinfo {pages} {89--92} (\bibinfo {year} {2011})}\BibitemShut {NoStop}%
\end{thebibliography}%

\appendix
\section{Device parameters}
\label{sec:SIparams}
\begin{figure}[h!]
\centering\includegraphics[]{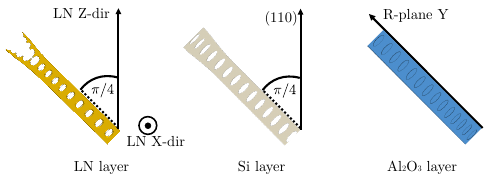}
\caption{\textbf{Orientations of the transducer beam with respect to the crystalline directions.} The LN and sapphire crystal orientations are as defined in the literature\cite{weis_lithium_1985,vodenitcharova_effect_2007}.} 
\label{fig:matFrames}
\end{figure}

In the following tables, Tabs. \ref{tab:paramOMC}, \ref{tab:paramEMC}, \ref{tab:paramN}, we provide the relevant parameters for the sOMC-based transducer. The crystal orientations of the lithium niobate and silicon layers are shown in Fig. \ref{fig:matFrames}.

\begin{table}[h!]
\caption{\textbf{Parameters of the OMC region.}  The silicon device layer is 220 nm thick.}
    \centering
\begin{ruledtabular}
\begin{tabular}{ c| c c c } 
    & Mirror & Defect & Partial mirror\\ \hline
   $a$  & $398$ nm & $193$ nm & $363$ nm \\
    $h_{x}$  & $197$ nm  & $98$ nm & $163$ nm\\
    $h_{y}$  & $530$ nm  & $429$ nm & $464$ nm\\
    $w$  & $700$ nm  & $700$ nm & $700$ nm\\
\end{tabular}
\end{ruledtabular}   
\label{tab:paramOMC}
\end{table}

\begin{table}[h!]
\caption{\textbf{Parameters of the EMC region.} The lithium niobate device layer is 100 nm thick. The tabulated values $h_x$, $h_y$, $w$ are given at the top of the lithium niobate layer. The ellipse values at the top of the silicon layer can be retrieved by calculating the sidewall width and allowing for a 50 nm misalignment buffer in the hole. The sidewall angles are 18° and 8° in the hole and at the beam edge respectively. The silicon beam is assumed to have 0° sidewall angle. For the taper-end cell the values are given for the top of the silicon layer. The polynomial $f$ is defined in the text.}
    \centering

\begin{ruledtabular}
\begin{tabular}{ c | c c c c } 
 & Taper-End (Si) & Taper-End  & Defect & Mirror  \\  \hline
   $a$  & $200$ nm & $200$ nm & $308$ nm & $450$ nm \\
    $h_{x}$   & $100$ nm & $f$  & $208$ nm & $245$ nm \\
    $h_{y}$  & $459$ nm & $675$ nm  & $321$ nm & $349$ nm   \\
    $w$  & $900$ nm &  $770$ nm  &$570$ nm & $570$ nm  \\
\end{tabular}   
\end{ruledtabular}
\label{tab:paramEMC}
\end{table}

\begin{table}[h!]
\caption{\textbf{Number of unit cells in each region} The table lists the number of transition cells between adjacent regions.}
    \centering
\begin{ruledtabular}
\begin{tabular}{ ccc } 
   Device region  & $N_{\rm uc}$  \\  \hline
    OMC mirror - OMC defect  & $10$  \\
   OMC defect - OMC partial mirror  & $10$  \\
   OMC partial mirror  & $8$  \\
   OMC partial mirror - taper end cell  & $4$  \\
   taper end cell - EMC defect & $4$   \\
   EMC defect - EMC mirror  & $5$  \\
\end{tabular} \\ 
\end{ruledtabular}
\label{tab:paramN}
\end{table}

To address the transducer's EMC we use 4 aluminum electrodes, i.e. 2 periods, 80 nm in width and 30 nm in thickness. The transducer design is straightforward to adapt to electrodes made of low-quasiparticle-lifetime and high-kinetic-inductance materials such as NbTiN.

The function describing how $h_x$ is transformed in the transition region is: $h_x(h_y)=d\cdot h_y^3+c\cdot h_y^2+b\cdot h_y+a$. The parameters listed in rising polynomial order are $[ -1271 \ \text{nm},\ 	9.365\ \text{nm}^{-1},\	-0.01875\ \text{nm}^{-2},\	1.238\cdot 10^{-5}\ \text{nm}^{-3}]$.

\section{Design considerations}
\subsection{Acoustic modes in the substrate}
\label{app:SAWs}
The choice of substrate is essential for the design of the release-free transducer. Acoustic excitations in the substrate such as surface or bulk acoustic waves (SAWs, BAWs) are the dominant channel of mechanical loss in our system. Surface acoustic waves in the substrate usually constitute the lowest frequency acoustic excitations at a given wavevector. Substrates with higher SAW velocity may therefore increase mechanical confinement in analogy to high refractive-index-contrast and optical confinement for optical devices.

\begin{figure}[ht!]
\centering\includegraphics[]{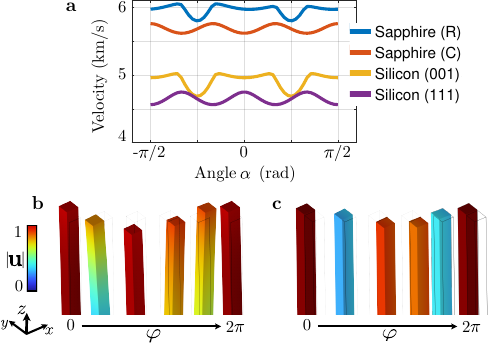}
\caption{\textbf{Acoustic properties of sapphire and silicon.} \textbf{a)} Sound velocities in a variety of substrates as a function of in-plane angle $\alpha$. We take the elasticity matrices from literature\cite{auld_acoustic_1990}. \textbf{b)-c)} Mode profiles as a function of phase for surface acoustic waves and transversal bulk acoustic waves respectively. The modes shown appear for R-plane sapphire at $\alpha=0$ and $\alpha=\pi/4$ respectively, with $k=0.8\cdot\frac{\pi}{400}\ \rm 1/nm$.}
\label{fig:saws}
\end{figure}

\begin{figure*}[ht!]
\centering\includegraphics[]{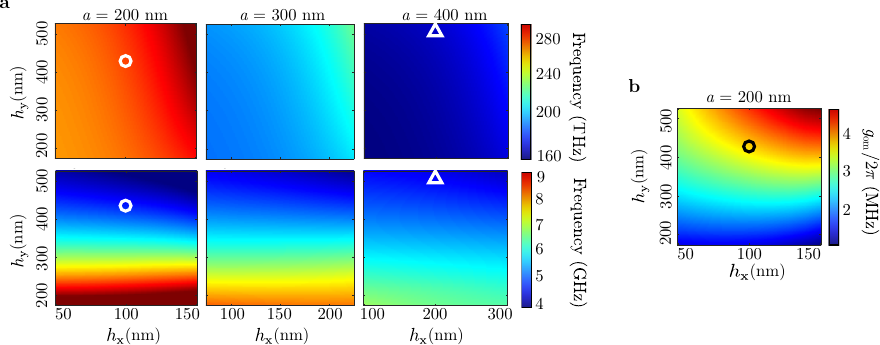}
\caption{\textbf{Geometric tuning of mode frequencies and unit-cell vacuum optomechanical coupling rate.} \textbf{a)} Unit cell optical (top) and mechanical (bottom) mode frequencies at the $X$-point as a function of ellipse parameters $h_x$, $h_y$. The three columns correspond to three different periods $a$. The approximate operating point of defect and mirror cell is marked with a white circle and triangle, respectively. \textbf{b)} Vacuum optomechanical coupling rate in an infinitely periodic $a=200\ \rm nm$ unit cell under counter-propagating phase-matching condition $k_m=2k_o=\pi/a$. The approximate operating point is marked with a black circle. } 
\label{fig:ucparams}
\end{figure*}
We study the acoustic velocities using a simple simulation model: we consider a thin (10 nm sides) and long beam of the substrate. Between the two $yz$-planes we impose a boundary condition $\mathbf{u}_1=\mathbf{u}_2 e^{-i\mathbf{k}(\mathbf{r}_1-\mathbf{r}_2)}$, with $\mathbf{k}=k\mathbf{e}_x$. We use the frequency of the lowest frequency mode found in this way to determine the sound velocity for a given $k$. The resulting velocities in silicon adequately reproduce the reported experimental results (Fig. \ref{fig:saws}a) \cite{coufal_precision_1994}. Depending on the in-plane rotation angle $\alpha$ of the crystal, we find Rayleigh-type surface acoustic waves or slow transverse bulk-type waves (Fig. \ref{fig:saws}b-c) \cite{tarasenko_theoretical_2021}.

As alluded to in main text, sapphire is an attractive substrate for several reasons including its high SAW velocity. However, even when the substrate's sound velocity is below that of silicon, such as for SiO$_2$ ($v\approx 3200\ \rm m/s$), the mechanical mode of the transducer can be confined by geometric softening \cite{kolvik_clamped_2023-1,sarabalis_release-free_2017}. Thus, the presented design principles apply to a wide variety of other thin-film and substrate choices.

\subsection{Optical and mechanical geometric frequency tuning}
\label{app:freqtuning}
We tune both the optical and mechanical mode frequencies in our design work, exploiting the mode frequencies' dependence on the unit-cell geometry. Considering a unit cell of a periodic waveguide, we show the $X$-point frequencies in Fig. \ref{fig:ucparams}a and the optomechanical coupling rates in Fig. \ref{fig:ucparams}b -- both as a function of unit cell parameters.
Because the optical field responds mostly to the effective index of the beam, there is no dramatic difference in scaling between $h_x$ and $h_y$. For the mechanical frequency, on the other hand, $h_y$, i.e. the dimension perpendicular to the beam, yields a much stronger effect compared to $h_x$. This asymmetry can be understood in the following way: increasing $h_y$ and decreasing $h_x$ both increases the effective mass. However, increasing $h_y$ softens the geometry and decreasing $h_x$ stiffens it. In the former case, the softening and the increase in mass act together to decrease the frequency, while in the latter case, the stiffening and the mass increase counteract. Optics and mechanics are affected quite differently by variations in $a$ and $h_y$ which allows us to create a partial mirror (Sec. \ref{sec:transition}).

In addition to mode frequencies, the shape of the band is also important. For example, a flat mechanical band, most prominent for unit cells with small $a$ and large $h_y$, is associated with lower group velocity and likely higher susceptibility to inhomogeneous broadening due to fabrication disorder \cite{safavi-naeini_two-dimensional_2014,wolff_brillouin_2016}. 
This creates a trade-off: small $a$ and large $h_y$ are desirable for high optomechanical coupling, but they also flatten the band and induce smaller feature sizes (Fig. \ref{fig:ucparams}).

\subsection{Optical and mechanical cavity spectra}
\label{app:cavSpectra}
\begin{figure*}[ht!]
\centering\includegraphics[]{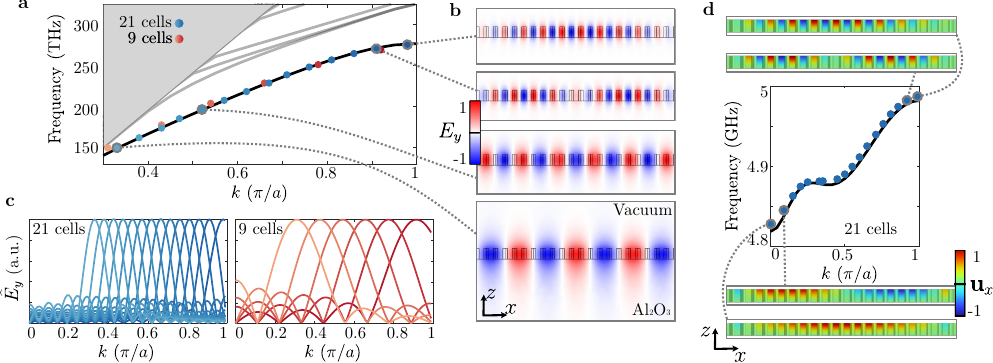}
\caption{\textbf{Mode spectra for optics and mechanics in an artificial cavity consisting only of defect unit cells and artificial mirrors.} \textbf{a)} Optical band diagram of the unit cell used in the cavity. The circles correspond to cavity modes. \textbf{b)} Selected mode profiles of cavity modes shown in cross section of the beam. \textbf{c)} Fourier transforms of the modes shown in a). \textbf{d)} same as a)-c) but for mechanics. In d) there is no substrate: the beam is fixed at the bottom and free on the sides.} 
\label{fig:fakecav}
\end{figure*}
To illustrate the optical spectra in the OMC we study a simplified toy model: instead of combining defect and mirror regions via an adiabatic transition region, the model consists of a series of identical defect cells terminated by a perfect electrical conductor boundary condition on both sides of the simulation region.  
Since we are not limited by the finite bandgap of the OMC in this case, we can now find all cavity modes below the light-line. Because the cells are identical, the cavity modes discretize the band structure associated with the infinite waveguide of the underlying cell.

Fig. \ref{fig:fakecav}a-b shows selected mode profiles along the band. The positions in $k$-space are found by Fourier transforming the field and agree with estimates based on the cavity length, i.e. $k_i=(N-i)/N\cdot\pi/a$, where $N$ is the number of cells of period $a$. Increasing the length of the cavity narrows the wavevector distribution, tightening demands on phase-matching. 
The mode profiles have a spatial modulation along the cavity length \cite{eichenfield_optomechanical_2009,chan_optical_2009-1}. At $k\approx \pi/a$ we have a single-peak, fundamental-like, mode. For lower $k$ the modulation exhibits an increasing number of peaks until a maximum at $k\approx \pi/(2a)$. Further decreasing $k$ decreases the number of peaks until we are back at the $\Gamma$-point with a fundamental-like mode. Low $k$ modes are obscured by the optical continuum.

Similarly, we now consider a mechanical toy model consisting of a series of identical defect cells terminated by a free boundary condition. To avoid phonon scattering into the continuum, we simulate the cavity as fixed at the bottom but without a substrate. Thus, we can trace the spatial modulation of the mechanical field from the $X$- to the $\Gamma$-point (Fig. \ref{fig:fakecav}d). The apparent spatial mismatch between the $X$-point mechanical mode and the $\pi/(2a)$ optical mode is also mentioned in the main text.

\subsection{Thin interfacial oxide layer }
\label{app:oxidelayer}
\begin{figure}[ht!]
\centering\includegraphics[]{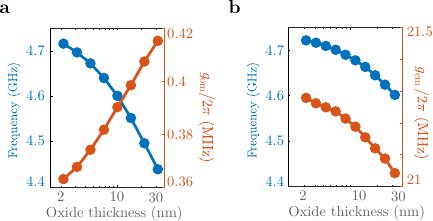}
\caption{\textbf{OMC and EMC frequency as a function of silicon dioxide layer thickness between the silicon and sapphire.} \textbf{a)} OMC mechanical frequency and optomechanical coupling rate. \textbf{b)} EMC mechanical frequency and electromechanical coupling rate.} 
\label{fig:oxsweep}
\end{figure}

In this section we briefly study the impact of a thin SiO$_2$ layer between sapphire and silicon. Fig. \ref{fig:oxsweep} shows the dependence of mechanical frequency and coupling on the silicon dioxide thickness. The coupling increases slightly as the layer thickens. Over the same thickness range, the optical frequency increases by around 0.5 THz, i.e. $\approx 5$ nm. 
Further increasing the oxide layer thickness eventually results in an SOI - based OMC like the one presented in \cite{kolvik_clamped_2023-1}. Both SOI and SOS allow for high radiation-limited optical quality factors. 

\subsection{Disorder studies}
\label{app:disorder}
\begin{figure}[h!]
\centering\includegraphics[]{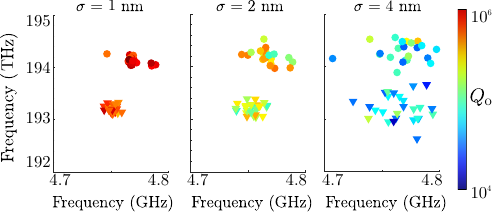}
\caption{\textbf{Disorder studies of sOMC and lOMC transducer.} We show supplementary data to Fig. \ref{fig:crossing}c in the main text. The same simulations as in main text are now plotted as a function of mechanical and optical frequencies. The colorscale denotes the optical quality factor. We plot the mode with largest $\gom$ from a set of 30 simulations for both the sOMC- (circles) and lOMC-based transducer (triangles).} 
\label{fig:SI_disorder}
\end{figure}

As described in the main text, we analyze the robustness of the design to artificially introduced disorder (Fig.\ref{fig:crossing}c-d). In addition to the data on coupling rates and mechanical quality factors shown in the main text, disorder also affects the mechanical and optical frequencies as well as the optical quality factor (Fig.\ref{fig:SI_disorder}). For disorder up to 4 nm, the optical wavelength varies on the scale of a few nanometers, while the mechanical frequency varies on the scale of $100 \ \rm MHz$. The optical quality factor is most affected by the disorder, dropping from $>10^6$ for the nominal design to an average of $2\cdot10^5$ at $\sigma=2\ \rm nm$.
We have performed similar studies of our silicon-on-insulator OMCs and compared them to experimental results. We conclude that our actual fabrication disorder yields similar $Q_o$ to those expected based on disorder simulations for around $\sigma=2-3\ \rm nm$.

\section{Calculation of coupling rates}
\subsection{Electromechanical coupling}
\label{app:gem}
In this section we provide an outline for the derivation of the electromechanical coupling rate $\gem$. The total energy of the piezoelectric oscillator (EMC section) consists of an electromagnetic, a mechanical, and a piezoelectric part:

\begin{equation}
\begin{aligned}
\label{eqn:H1}
    \mathcal{H}=&\frac{1}{2}\int \epsilon |\mathbf{E}|^2+\mu |\mathbf{H}|^2 \ d\mathbf{r} +\\
    &+\frac{1}{2}\int \rho |\dot{\mathbf{V}}|^2+S_{ij}c_{ijkl}S_{kl} \ d\mathbf{r} +\\
    &- \int \mathbf{D_m}\cdot \mathbf{E} \ d\mathbf{r},    
\end{aligned}
\end{equation}
where $\mathbf{V}$ is the displacement field and $\mathbf{D_m}$ is the electrical electrical displacement field resulting from mechanical motion through the piezoelectric effect. The other quantities follow common convention.
For simplicity we focus on only one mode at frequency $\omega$. Then the kinetic energy can be rewritten in terms of the displacement: $|\dot{\mathbf{V}}|=\omega^2|\mathbf{U}|$. 
We now quantize the fields as:
\begin{equation}
\begin{aligned}
\label{eqn:gem_quantization}
    \mathbf{E}&=\mathbf{e}\hat{c}+h.c.\\
    \mathbf{U}&=\mathbf{u}\hat{b}+h.c.\\
    \mathbf{D_m}&=\mathbf{d_m}\hat{b}+h.c.,    
\end{aligned}
\end{equation}
where $\hat{c}$ and $\hat{b}$ are the ladder operators for the microwave and mechanical modes respectively. As will be shown below, using this expansion the Hamiltonian $\mathcal{H}$ can be written as
\begin{equation}
    \mathcal{H}=\hbar\om b^\dagger b+\hbar\omega_\mu c^\dagger c + + \hbar \gem (b c^\dagger +b^\dagger c)
\end{equation}
in the rotating-wave approximation and with appropriate field normalizations. Next, we find the appropriate normalization of the simulated mode profiles. We start with the electromagnetic component:

\begin{equation}
    \begin{aligned}
        \frac{1}{2}\int \epsilon |\mathbf{E}|^2+\mu |\mathbf{H}|^2 \ d\mathbf{r}
        &= \Big(\int \epsilon |\mathbf{e}|^2+\mu |\mathbf{h}|^2 \ d\mathbf{r}\Big) c^\dagger c \\
        &= \Big(2\int \epsilon |\mathbf{e}|^2\ d\mathbf{r}\Big)c^\dagger c \\
        &\equiv \hbar \omega_\mu c^\dagger c,
    \end{aligned}
\end{equation}
where we have used the commutator $[c,c^\dagger]=1$ and $\int \epsilon |\mathbf{e}|^2d\mathbf{r}=\int \mu |\mathbf{h}|^2d\mathbf{r}$. The same procedure can be applied to the mechanical energy. Hence, we arrive at the normalizations:
\begin{equation}
\begin{aligned}
    &U_\mu=2\int \epsilon |\mathbf{e}|^2\ d\mathbf{r}={\hbar\omega_\mu} \\ 
    &\Um=2\om^2\int \rho |\mathbf{u}|^2\ d\mathbf{r}={\hbar\om}    
\end{aligned}
\end{equation}
These normalizations connect the fields $\mathbf{u}$, $\mathbf{e}$, $\mathbf{d_m}$ to the simulated fields $\mathbf{u}'$, $\mathbf{e}'$, $\mathbf{d'_m}$:
\begin{equation}
\begin{aligned}
\label{eqn:gem_norm}
    \mathbf{u}&=\sqrt{\frac{\hbar\om}{\Umprime}}\mathbf{u'}\\
    \mathbf{e}&=\sqrt{\frac{\hbar\omega_\mu}{U'_\mu}}\mathbf{e'}\\
    \mathbf{d_m}&=\sqrt{\frac{\hbar\om}{\Umprime}}\mathbf{d'_m},    
\end{aligned}
\end{equation}
where we use the mechanical and electrical energies of the mode fields:
\begin{equation}
    U'_\mu=2\int \epsilon |\mathbf{e'}|^2\ d\mathbf{r} \quad \text{and} \quad \Umprime=2\om^2\int \rho |\mathbf{u'}|^2\ d\mathbf{r}
\end{equation}
Using \ref{eqn:gem_quantization} we examine the interaction part of the Hamiltonian \ref{eqn:H1} neglecting the fast-varying terms in $ bc $ and $b^\dagger c^\dagger $: 
\begin{equation}
\begin{aligned}
    \mathcal{H}_{\rm int}=-\int \mathbf{D_m}\cdot\mathbf{E} \ d\mathbf{r} &= \hbar ( \gemtilde b c^\dagger + \gemtilde^\star b^\dagger c)  \\
\end{aligned}
\end{equation}
with $\hbar\gemtilde = -\int \mathbf{d_m}\cdot\mathbf{e}^\star d\mathbf{r}$. We absorb the phase of $\gemtilde$ into the ladder operators such that the coupling rate is a real and positive quantity without loss of generality. That is, we take the absolute value of $\gemtilde$, i.e. $\left\|\gemtilde\right\|\equiv \gem$. Inserting the field normalizations \ref{eqn:gem_norm}, we arrive at:
\begin{equation}
\label{eqn:gem_final}
\begin{aligned}
    \mathcal{H}_{\rm int}=-\int \mathbf{D_m}\cdot\mathbf{E} &= \hbar \gem (b c^\dagger +b^\dagger c) \\
    \text{with } \qquad \gem&\equiv \frac{\sqrt{\omega_\mu \om}}{\sqrt{\Umprime U'_\mu}}\left\|\int \mathbf{d'_m}\cdot\mathbf{e'}^\star  \ d\mathbf{r}\right\|
\end{aligned}
\end{equation}
This expression holds for an oscillating electric field $\mathbf{e'}$. We retrieve instead a static electric field $\mathbf{e'}$ from a stationary simulation such that $\mathbf{e'}=\mathbf{e'}^\star$. We re-write the electromagnetic energy $U'_\mu$ associated with the oscillating field in terms of the static capacitance $C_{\rm IDT}+C_\mu$ associated with the inter-digitated electrodes and the microwave resonator or qubit  as $ U'_\mu=2\int \epsilon |\mathbf{e'}|^2\ d\mathbf{r}=4U_{\rm electrostatic}=2(C_{\rm IDT}+C_\mu )V_0^2$ with $V_0$ the voltage applied in the static simulation. Our expression for the piezomechanical coupling rate $\gem$ agrees with the expressions derived by Bl\'{e}sin et al.\cite{blesin_quantum_2021} and is similar to the expression used in Chiappina et al.\cite{chiappina_design_2023-1}. It also numerically agrees with the $\gem$ results we obtain through Foster synthesis following the method in Arrangoiz-Arriola et al.\cite{arrangoiz-arriola_engineering_2016} for the presented release-free EMC and a suspended phononic crystal defect\cite{arrangoiz-arriola_resolving_2019}.  The Foster synthesis method requires more simulation time and is more sensitive to mode crowding, so we primarily use the overlap integral approach described above.

The electromechanical coupling rate at fixed frequency $\omega_\mu$ is proportional to both the zero-point mechanical motion and the zero-point voltage. The latter scales as the square root of the microwave impedance $Z_\mu^{1/2}$ at fixed frequency: $\gem \propto \omega_\mu/\sqrt{ U'_\mu} = Z_\mu^{1/2}$ (eqn. \ref{eqn:gem_final}) when keeping the frequencies fixed. In order to compare our results with those of prior work, we take $C_\mu=70\ \rm fF$ throughout the main text\cite{banker_photonic_nodate,chiappina_design_2023-1} without loss of generality. This capacitance $C_\mu=70\ \rm fF$ corresponds to an impedance of $Z_\mu=\frac{1}{\omega_\mu (C_{\rm IDT}+C_\mu)}\approx 474 \ \Omega$, whereas high-impedance resonators with $Z\gg 1\ k \Omega$ have been demonstrated\cite{niepce_high_2019,meesala_non-classical_2024}.

LNs dielectric loss tangent\cite{zorzetti_millikelvin_2023}, $ \tan \delta_{\rm LN}=1.7\cdot10^{-5}$, is several orders of magnitude greater than the loss tangents of sapphire and silicon \cite{luiten_ultrahigh_1993,martinis_ucsb_2014}. The microwave loss rate contributed by the LN is set by the participation ratio of the microwave field in the piezoelectric material $\kappa_{\mu,\rm LN}=\frac{C_{\rm IDT}}{C_{\rm IDT}+C_\mu} \omega_\mu\tan \delta_{\rm LN}$ \cite{chiappina_design_2023-1}.
For the parameters in this work we have $\kappa_{\mu,\rm LN}/2\pi=340\ \rm Hz$, which is of similar magnitude as state-of-the-art transmon qubits on sapphire\cite{place_new_2021} with $\kappa_{\mu,i}/2\pi\approx500\ \rm Hz$. In the long-run, this may provide motivation to further reduce the static capacitance $C_{\rm IDT}$. The microwave piezoelectric loss is unlikely to limit the performance in near-term experiments. It is well below the losses associated with kinetic-inductance resonators \cite{meesala_non-classical_2024}.

\begin{figure}[ht!]
\centering\includegraphics[]{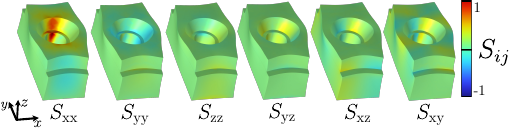}
\caption{\textbf{Mechanical mode in the EMC defect cell.} The plot series corresponds to different components of the strain tensor showing that the $S_{\rm xx}$ component is strongest.} 
\label{fig:strainplots}
\end{figure}

The mechanical fields are extracted by simulating the device's eigenmodes. In this simulation, the electrodes present a metallic boundary condition at their interfaces. The magnitude of the applied potential does not affect the eigenmode solution. To illustrate the piezomechanical coupling mechanism, we plot the strain profile in the EMC unit cell, see Fig. \ref{fig:strainplots}. As described in the main text, the $S_{\rm xx}$ component is strongest, which is exploited by $E_{\rm x}$ and the associated component the piezoelectric coupling matrix, $e_{11}$.

\subsection{Optomechanical coupling}
\label{sec:SIgom}

We calculate the optomechanical coupling using an overlap integral of optical and mechanical fields. We take into account the moving-boundary contributions between silicon and air -- which dominate -- as well as between silicon and sapphire. We also consider the bulk photoelasticity in silicon and sapphire \cite{yariv_optical_2003}. The bulk photoeleasticity contribution amounts to $\approx 4\%$ of the total coupling rate for our mechanical mode.

As we explore below, the unconventional operating point chosen in this work with $\km=\pi/a$ and $\ko=\pi/(2a)$ relies on counter-propagating optical waves to generate optomechanical interactions adding up from one unit cell to the next\cite{kolvik_clamped_2023-1}. In effect, we use backward Brillouin interactions in an optomechanical crystal cavity \cite{schmidt_convergence_2022}.
In the standing-wave cavity, where photons and phonons naturally travel in both directions, the overlap integral takes the same form as for the co-propagating scenario using $\Gamma$-point mechanical modes with near-vanishing longitudinal wavevector \cite{chan_laser_nodate}. For example, the moving-boundary term is still given by\cite{wolff_chapter_2022,johnson_perturbation_2002}:
\begin{equation}
\label{eqn:g_rp}
    \gom|_{\mathrm{m.b.}}=\sqrt{\frac{\hbar}{2 \omega_{\mathrm{m}}}} \frac{\omega_{\mathrm{o}}}{2} \frac{\int(\boldsymbol{u}(\boldsymbol{r}) \cdot \boldsymbol{n})(\Delta \epsilon|\boldsymbol{E}^{\|}|^2-\Delta(\epsilon^{-1})|\boldsymbol{D}^{\perp}|^2) \mathrm{d} A}{\sqrt{\int \rho|\boldsymbol{u}(\boldsymbol{r})|^2 \mathrm{~d}^3 \boldsymbol{r}} \int \epsilon(\boldsymbol{r})|\boldsymbol{E}(\boldsymbol{r})|^2 \mathrm{~d}^3 \boldsymbol{r}},
\end{equation}
with the fields having forward and backward components, e.g.: $\boldsymbol{E}\propto \tilde{\boldsymbol{E}}_{\mathrm{f}} e^{i x k_{\mathrm{o}, \mathrm{f}}}+\tilde{\boldsymbol{E}}_{\mathrm{b}} e^{i x k_{\mathrm{o}, \mathrm{b}}}$. For the unit-cell coupling in a periodic waveguide (see Fig. \ref{fig:ucparams}b) where the pump and the Stokes photons are counter-propagating, we use a modified  expression \cite{qiu_stimulated_2013,wolff_chapter_2022}. The optical fields appearing in the numerator of eqn. \ref{eqn:g_rp} then become: 
$|\boldsymbol{E}^{\|}|^2\  \rightarrow \ (\boldsymbol{E}^{\|})^2$, 
and the same for $\boldsymbol{D}^{\perp}$. This corresponds to changing $k \ \rightarrow \ -k$ for one of the two participating optical fields. We test our simulation code by calculating the optomechanical coupling rates for (1) the traveling-wave silicon nanowires demomstrated in Qiu et al. \cite{qiu_stimulated_2012} and Van Laer et al.\cite{van_laer_interaction_2015} for both co- and counter-propagating cases, and (2) the suspended OMC cavity demonstrated in Chan et al.\cite{chan_laser_2011}. We find good agreement for both the magnitude and the relative sign of the moving-boundary and photoelastic contributions in the traveling-wave silicon nanowires \cite{qiu_stimulated_2012, van_laer_interaction_2015}. In those cases, we convert between Brillouin gain coefficient and vacuum optomechanical coupling rate \cite{van_laer_unifying_2016}. For the suspended OMC cavity, we find a coupling rate of $\gomopi\approx 1 \ \text{MHz}$, with constructively interfering moving-boundary and photoelastic contributions. In addition, our simulations are in good agreement with our measurements for a release-free optomechanical crystal on a silicon dioxide substrate \cite{kolvik_clamped_2023-1}.

\section{Simulation techniques}
\subsection{Meshing in finite element simulations}
\begin{figure}[ht!]
\centering\includegraphics[]{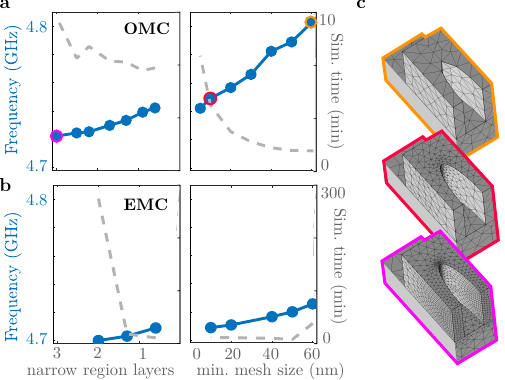}
\caption{\textbf{Mechanical frequency as a function of mesh resolution.} The dashed gray line denotes the simulation time. \textbf{a)} An OMC is simulated with a symmetry plane along its long axis. We sweep both the minimum mesh size (right) and then also the resolution in narrow regions (left). \textbf{b)} Same as a) but for the an EMC cavity and without symmetry planes. \textbf{c)} Mesh of unit cells in the OMC along the curve in (a). } 
\label{fig:meshsweep}
\end{figure}

Achieving accurate simulations requires sufficiently high mesh resolution. However, a higher mesh resolution also increases the simulation time such that a good balance must be found. Because of the variety of crystalline materials involved we cannot easily invoke symmetries to reduce the system's degrees of freedom. As our design critically depends on mechanical frequency matching between subsections, we use the mechanical frequency as our key parameter setting this balance between mesh accuracy and simulation time. We examine the mechanical frequency of a pure OMC as a function of the mesh resolution (Fig. \ref{fig:meshsweep}a). 
We find that the mechanical frequency decreases monotonically with increasing mesh resolution. The EMC frequency is affected less by mesh refinement (Fig. \ref{fig:meshsweep}b). Coupling rates and optical frequencies are practically unaffected over this mesh range. The simulation time on the other hand is generally much longer for simulations involving piezoelectricity. 
The mesh used for the simulations in the bulk of this work corresponds to approximately 20 nm in minimum mesh size in Fig. \ref{fig:meshsweep}a.
In our case, the simulation times for a single full transducer, including both optical and piezomechanical parts along with coupling calculations, can be up to one hour.

\subsection{Perfectly matched layer}
The results of the simulations depend on the boundary conditions at the edges of the simulation domain. We employ a perfectly matched layer (PML) at the edges to attenuate incoming optical and mechanical fields and create an effectively boundless domain. We use an exponential function for the attenuation factor: 
\begin{equation}
     f_{\rm pml}(r)=A\Big(\exp\Big(\frac{r-R_{\rm start}}{R_{\rm 0}}\Big)-1\Big) \quad \text{for } \quad r>R_{\rm start} 
\end{equation} 

\begin{figure*}[ht!]
\centering\includegraphics[]{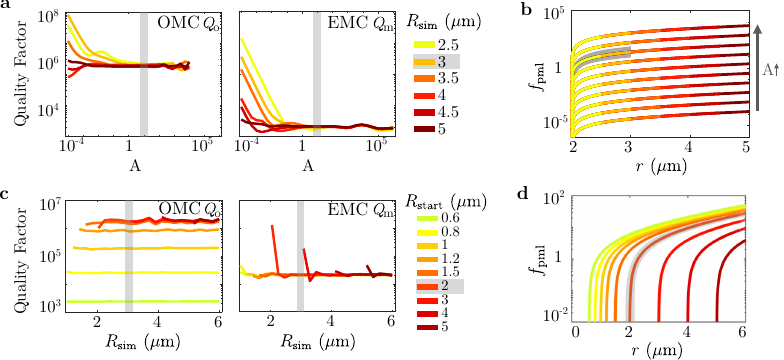}
\caption{\textbf{Effect of perfectly matched layer (PML) properties on the radiation-limited quality factors.} For the short optomechanical crystal (sOMC): \textbf{a)} Optical (left) and mechanical (right) quality factor as a function of PML amplitude $A$. The colors correspond to varying simulation domain sizes, $R_{\rm sim}$. The shaded regions indicate the operating point in this work. \textbf{b)} PML function $ f_{\rm pml}(r)$ as a function of $A$. The color coding is the same as in a). \textbf{c-d)} Same as \textbf{a-b)} but for different PML parameters. } 
\label{fig:pml}
\end{figure*}

where $A$ is the amplitude of the PML. It begins to act at $R_{\rm start}$ with a characteristic slope related to $R_{\rm 0}$. Then it increases until the end of the simulation domain is reached, $R_{\rm sim}$. To attenuate optical (mechanical) fields, we use $ f_{\rm pml}$  as imaginary part of the refractive index (density). The overall scale is determined by the real part of refractive index (density).

In Fig. \ref{fig:pml} we examine the behavior of the PML. We aim to operate in a parameter region where $Q$ is stable to perturbations, i.e. a plateau. Away from the plateau $Q$ fluctuates and becomes inconsistent (Fig. \ref{fig:pml}a). We find a larger spread in $\Qm$ compared to $\Qo$ in their plateau regions.
When $R_{\rm start}$ is too small, the PML may absorb significant parts of the evanescent field of the mode which is undesirable. We find that starting the PML at $R_{\rm start}=2 \ \rm \mu m$ the quality factors converge. Although the initial rise of $f_{\rm pml}$ looks fast on logarithmic scale (Fig. \ref{fig:pml}b,d) the exponential shape of $ f_{\rm pml}$ ensures that it rises from zero at $R_{\rm start}$ so that there are no reflections because of an abrupt change in the material.

\subsection{Design optimization}
\begin{figure}[ht!]
\centering\includegraphics[]{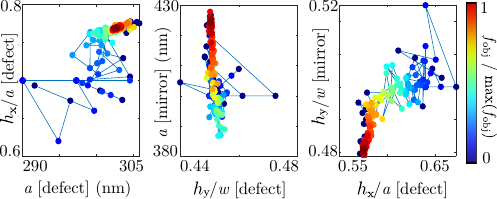}
\caption{\textbf{Optimization of design parameters.} We use a Nelder-Mead algorithm to optimize pure EMC parameters for the local maximum of the objective function $f_{\rm obj}=Q_m \gem$. Each dot corresponds to a simulation, i.e. a step in the algorithm. Similar optimizations are run for the pure OMC with objective function X, and for the full transducer with objective function Y.} 
\label{fig:optim}
\end{figure}
\label{app:optim}
In our design process we use a Nelder-Mead algorithm to optimize the key design parameters\cite{chan_laser_nodate}. The most important optimizations are those of the pure OMC and EMC. In those, we aim for the highest coupling rate while achieving a minimum value for the quality factor (Fig. \ref{fig:optim}). For every iteration, we also include limits on minimum feature size, bandgap size and mechanical and optical frequency.

As described in the main text, introduction of the partial mirror changes the fields in the OMC (Fig. \ref{fig:partialMirrorsweep}). To quickly examine this type structure we simulate the OMC with the partial mirror on one side followed by a transition into the nominal OMC mirror. This simulation yields $\gom$ while a second simulation without the OMC nominal mirror yields $Q_o$. This can speed up optimizations because these simulations tend to be significantly faster than those that include piezoelectricity.

Lastly, we optimize the final design of the full transducer. Because of the long simulation times, the number of random initial points and the number of iterations are both limited. Thus, we focus on the design parameters that are most critical for the hybridization, i.e. the partial mirror and the EMC. To also factor in the robustness of the device, we let each optimization step consist of two simulations separated by 2\% in EMC period. By using only the result with inferior figure of merit, we encourage the algorithm to decrease the sensitivity of the design to geometric perturbations. For the presented sOMC and lOMC transducers, the optimization maximizes the objective function $f_{\rm obj}=\gom\gem\Qo\Qm$, where $\Qo{_{,\rm max}}=10^6$ and $\Qm{_{,\rm max}}=10^4$. 
In future work, it might be interesting to introduce a maximum value for $\gem$ such that more mechanical energy can be devoted to the optomechanical region that is expected to limit performance. With sufficient closed-form knowledge of the thermal and added noise behavior of the device, eventually the full efficiency-bandwidth product (Sec.\ref{Sec:stateOfTheArt}) or other protocol-specific metrics could serve as an objective function to better distribute the performance between the various sections of the transducer.

\section{Overview of relevant literature}
\begin{table*}[ht!]
\caption{\label{tab:table3} \textbf{Overview of simulated metrics of short and long release-free transducers, and comparison to experimental and simulated (sim.) state-of-the-art suspended implementations.} For the simulated works we have assumed the same extrinsic and intrinsic optical and microwave loss rates as in Meesala et al.\cite{meesala_non-classical_2024}. We estimate $\Equbit$ for future 2D OMC-based transducers by assuming half the reported $\gom$ of the pure 2-D OMCs, $\etaem=0.95$, and microwave loss rates as in Meesala et al.\cite{meesala_non-classical_2024}. We welcome feedback from the authors of the relevant articles in case we have misinterpreted their results.}
\begin{ruledtabular}
\begin{tabular}{ l | c c c c c }
    & Type & $\gomopi$  & $\gemopi$  &  $\Cem$ (est.)& $\Equbit$ (est.)  \\ 
    &   & (MHz) & (MHz)  &  (-) &  (pJ/qubit)  \\ \hline
    MacCabe et al. \cite{maccabe_nano-acoustic_2020}            & 1-D OMC  & 1.15 & - & - & -   \\
    Mirhosseini et al. \cite{mirhosseini_superconducting_2020}     & 1-D & 0.43 & 2.25 & 112 & 1.8   \\
     Jiang et al. \cite{jiang_optically_2023}                   & 1-D &  0.41 & 0.42 & 0.22 & 1.3   \\
     Weaver et al. \cite{weaver_integrated_2024}     & 1-D & 0.53 & 7.4 & 24 & 1.7  \\
     Meesala et al. \cite{meesala_non-classical_2024}     & 1-D & 0.27 & 1.2 & 22 & 1.1 \\
    Chiappina et al. (sim.) \cite{chiappina_design_2023-1}     & 1-D & 0.83 & 2.8 & 119 &0.12   \\
    Zhao et al. \cite{zhao_quantum-enabled_2024}     & 1-D & 0.34 & 0.19 & 29 & 0.33 \\
    Ren et al.  \cite{ren_two-dimensional_2020}     & 2-D OMC & 1.06 & -  & - & (0.11)    \\
    Mayor et al.  \cite{mayor_two-dimensional_2024}    & 2-D OMC & 0.86 & - & - & (0.29)  \\
    Kolvik et al. \cite{kolvik_clamped_2023-1}     & 1-D release-free OMC  & 0.5 & - & - & -   \\
    \textbf{This work: sOMC} (sim.)    & 1-D release-free & 0.24  & 6.3 & 49 & 1.4   \\ 
    \textbf{This work: lOMC} (sim.)    & 1-D release-free & 0.34  & 2.4 & 26 & 0.71  \\
\end{tabular}
\end{ruledtabular}
\label{tab:comparison}
\end{table*}
Here, we provide a non-exhaustive overview of related piezo-optomechanics transducers to better place our design efforts in the context of prior work. We list both experimental and simulation results. Several relevant parameters are not explicitly available in the literature, so we calculate them ourselves. In some cases, these calculations have large uncertainty and should be considered as rough estimates (est.). We have not added information on the disorder sensitivity (Secs. \ref{sec:robustness}, \ref{app:disorder}) to the table due to lack of information in the literature.

\end{document}